\documentclass[a4paper,fleqn,usenatbib,useAMS]{mnras}

\usepackage[T1]{fontenc}
\usepackage{ae,aecompl}

\usepackage{graphicx}	
\usepackage{amsmath}	
\usepackage{amssymb}	
\usepackage{bm}		
\usepackage{pdflscape}	

\usepackage{ulem}
\usepackage{times}
\usepackage{txfonts}
\usepackage{color}

\newcommand{\fracb}[2]{\left(\frac{#1}{#2}\right)}

\usepackage[dvipsnames]{xcolor}
\definecolor{blazeorange}{rgb}{1.0, 0.4, 0.0}
\definecolor{seagreen}{rgb}{0.18, 0.55, 0.34}
\definecolor{rufous}{rgb}{0.66, 0.11, 0.03}
\definecolor{royalfuchsia}{rgb}{0.79, 0.17, 0.57}
\definecolor{scarlet}{rgb}{1.0, 0.13, 0.0}
\definecolor{royalpurple}{rgb}{0.47, 0.32, 0.66}
\def\heasoft{\hbox{\rm{\small HEASOFT}}}
\def\nustardas{\rm {\small NUSTARDAS}}

\title[Disk-Corona Model \& Mass Estimates of Hol IX X-1]
{The Disk-Corona Model and Mass Estimates of the Ultraluminous X-ray Source Holmberg IX X-1}

\author[]{
Ramandeep Gill,$^{1,2,3}$\thanks{E-mail: rsgill.rg@gmail.com},
Eda Sonbas,$^{4,1}$
Kalvir S. Dhuga,$^{1}$
and Ersin G\"o\u{g}\"u\c{s}$^{5}$
\\
$^{1}$Department of Physics, The George Washington University, Washington, DC 20052, USA\\
$^{2}$Department of Natural Sciences, The Open University of Israel, P.O Box 808, Ra'anana 43537, Israel\\
$^{3}$Astrophysics Research Center of the Open university (ARCO), The Open University of Israel, P.O Box 808, Ra'anana 43537, Israel\\
$^{4}$Adiyaman University, Department of Physics, 02040 Adiyaman, Turkey\\
$^{5}$Faculty of Engineering and Natural Sciences, Sabanc\i~University, Orhanl\i~- Tuzla, Istanbul 34956, Turkey
 }

\date{Accepted XXX. Received YYY; in original form ZZZ}

\pubyear{2020}

\date{Received xxx; accepted xxx}
\begin{document}
\label{firstpage}
\pagerange{\pageref{firstpage}--\pageref{lastpage}}
\maketitle

\begin{abstract}
The origin of the variable X-ray emission in the $(0.3-30)\,$keV energy range of ultraluminous X-ray sources (ULXs) remains unclear,  
making it difficult to constrain the mass of the central compact object. Many ULXs show X-ray luminosities, $L_X>10^{39}\,{\rm erg\,s}^{-1}$, 
that exceed the Eddington limit ($L_{\rm Edd}$) of stellar-mass black holes (BHs) commonly found in Galactic BH binaries. 
Sub-critical accretion ($L<L_{\rm Edd}$) on to an intermediate-mass BH is one attractive 
scenario with the alternative being super-critical accretion on to a stellar-mass BH. Broadband X-ray emission in the former scenario can be 
explained using the canonical disk plus Comptonizing corona model, whereas in the latter scenario radiation pressure driven massive winds 
lead to complex spectra that are inclination angle dependent. Here we fit the broadband (optical/UV to X-ray) spectrum of the persistently 
bright ULX Holmberg IX X-1 with the disk-corona plus irradiated outer disk model in an effort to constrain the BH mass. We 
use a one-zone time-dependent numerical code to exactly solve for the steady-state properties of the optically thick coronal photon-electron-positron 
plasma. Our modelling suggests that Holmberg IX X-1 hosts a stellar mass BH, with mass $4\lesssim(\hat M_{\rm BH}\equiv\alpha M_{\rm BH}/M_\odot)\lesssim10$ 
where $1/6\leq\alpha<1$ for a spinning (Kerr) BH, undergoing super-critical accretion ($L_{\rm Bol}/L_{\rm Edd}\sim20\alpha$). In our model, the X-ray 
spectrum below $10\,$keV is explained with an absorbed multi-colour disk spectrum having inner disk temperature $k_BT_{\rm in}\sim(2.2-2.9)\,$keV. 
An additional cooler thermal spectral component, as found in many works and not included in our modeling, is required. The hard excess above $10\,$keV, 
as seen by \textit{NuSTAR}, arises in a photon-rich optically-thick Comptonizing spherical corona with optical depth $\tau_T\sim3.5$ and particle 
temperature $k_BT_e\sim14\,$keV. 
\end{abstract}

\section{Introduction}
Ultraluminous X-ray sources (ULXs) are non-nuclear point sources found in nearby galaxies with X-ray luminosities,  
$L_X>L_{\rm Edd}\approx1.3\times10^{39}(M_{\rm BH}/10M_\odot)\,{\rm erg\,s}^{-1}$, in excess of the Eddington luminosity 
of a $10M_\odot$ stellar-mass black 
hole (BH) that is routinely observed in Galactic BH binaries \citep[GBHB; see, e.g., reviews by][]{Feng-Soria-11,Kaaret+17}. 
Owing to their flux variability on timescales of hours to years, the majority of ULXs are thought to be powered by accretion 
on to a compact object. At least six of them that showed X-ray pulsations were identified with having a neutron star (NS) accretor \citep{Bachetti+14,Furst+16,Israel+17a,Israel+17b,Carpano+18,Sathyaprakash+19,Rodriguez-Castillo+20}. For the large number of remaining sources 
that do not show any pulsations, the mass of the compact object is a matter of intense debate. 
Two plausible scenarios have been offered to explain their super-Eddington luminosities: (i) Many of 
the more luminous sources with $L_X > 10^{40}\,{\rm erg\,s}^{-1}$ must be powered by sub-critical accretion on to an 
intermediate-mass black hole (IMBH) with $M_{\rm BH}\sim(10^2-10^4)M_\odot$ 
\citep[e.g.,][]{Colbert-Mushotzky-99,Makishima+00,Miller-Colbert-04,Miller+04,Farrell+09}, 
which can accommodate super-Eddington luminosities, and (ii) lower luminosity, but still super-Eddington, sources can be 
explained with super-critical accretion \citep[e.g.][]{Begelman-02,Poutanen+07,Dotan-Shaviv-11} on to a stellar-mass BH with $M_{\rm BH}\sim(10-10^2)\,M_\odot$, 
in which a strong outflow can further boost the isotropic-equivalent luminosity by relativistic beaming \citep{Ohsuga-Mineshige-11}. 
However, optical observations of extended highly ionized nebulae surrounding extremely luminous sources, e.g. Holmberg IX X-1, 
have placed strong constraints on relativistic beaming and instead favour isotropic emission \citep[][]{Moon+11}.

The highly luminous ULXs typically display two spectral components in the $(0.3-10)\,$keV energy range \citep[e.g.,][]{Miller+03,Vierdayanti+10,Miller+13}, 
comprising of a softer quasi-thermal excess below $2\,$keV and a harder power-law component above that energy, where the latter 
also shows spectral curvature in the $\sim(3-10)\,$keV energy range \citep[e.g.,][]{Feng-Kaaret-05,Stobbart+06,Gladstone+09,Walton+11}. 
Apart from the spectral curvature, these are analogous to the two-component spectra of GBHBs. In these sources, the softer component 
is modeled as multi-color disk (MCD) blackbody emission originating in the inner accretion disk, with characteristic temperature $k_BT_{\rm in}\sim1\,$keV, 
where $k_B$ is the Boltzmann constant, and the harder power-law is thought to arise in an optically thin (Thomson optical depth $\tau_T\lesssim1$) 
Comptonizing corona \citep[see][for a review]{Remillard-McClintock-06}. When the same model is applied to ULXs it yields low disk temperatures 
with $k_BT_{\rm in}\sim(0.1-0.3)\,$ keV \citep[e.g.,][]{Miller+03,Cropper+04} and an optically thick ($\tau_T>6$) corona with 
electron temperature $k_BT_e\sim(1-2)\,$keV. In a \citet{Shakura-Sunyaev-73} accretion disk, the canonical model used to describe GBHBs, 
the inner disk temperature scaling $T_{\rm in}\propto M_{\rm BH}^{-1/4}$ implies that ULXs 
are indeed powered by IMBHs. One of the main hurdles for this interpretation is that it requires an unreasonably high mass being 
held in stellar clusters to explain the production efficiency of ULXs \citep{King-04}.

The spectral curvature revealed by \textit{XMM-Newton} in the $\sim(3-10)\,$keV energy range was confirmed, as well as extended to 
energies above $10\,$keV, by \textit{NuSTAR} observations that also revealed a spectral turnover \citep[e.g.][]{Bachetti+13,Walton+14}. 
These observations further revealed the 
necessity of an additional spectral component above $10\,$keV, often assumed to be a cutoff power law, when modelling the broadband spectra 
of several ULXs. Therefore, any continuum model that shows an exponential drop in flux density, namely a Wien spectral tail, at high 
energies has been found to be inconsistent with \textit{NuSTAR} data of many ULXs \citep[e.g][]{Mukherjee+15,Walton+15,Luangtip+16,Walton+17}. 
In addition, this extra spectral component has also been identified in ULXs with a bonafide NS accretor, thus providing tantalizing hints that 
other non-pulsating ULXs may also host NSs rather than BHs \citep[][]{Pintore+17,Koliopanos+17,Walton+18}.

To understand the origin of the broadband X-ray spectra of ULXs as well as characterize the spectral variability a variety of empirical 
models are used \citep[e.g.,][]{Stobbart+06}. The best fits to X-ray data below $\sim10\,$keV are provided by dual thermal models that feature some combination of 
pure blackbody and an MCD blackbody spectra \citep{Mitsuda+84}. Equally good or better fits are provided by physically 
motivated models featuring an accretion disk \citep[e.g., DISKPN][]{Gierlinski+99} plus a Comptonizing corona spectrum obtained from 
self-consistent models, e.g. compTT \citep{Titarchuk-94} and EQPAIR \citep{Coppi-00}. In many cases in which \textit{NuSTAR} observations 
are used, the thermal models are supplemented by a cutoff 
power law model CUTOFFPL or a simple Comptonization model SIMPL \citep{Steiner+09} to explain the hard excesses above $10\,$keV. All of these 
models constrain the properties of the accretion disk and the steady state corona, however, they do not provide any direct constraints on the 
mass of the compact object apart from the normalization of the disk component that does depend on $M_{\rm BH}$.

In this work, we remedy this shortcoming by incorporating the mass of the accretor, assumed here to be a BH of mass $M_{\rm BH}$, 
into the model parameters, yielding 
a more complex behavior, where it not only affects the overall normalization but also the spectral shape, thus making it more self-consistent. 
The underlying model builds on the successes of the two-component disk plus Comptonizing corona model, and just like EQPAIR, it allows for 
hybrid particle distributions, such as thermal plus non-thermal. However, being a time-dependent one-zone model it evolves the properties 
of the corona until it reaches a steady state and yields parameters like the Thomson optical depth ($\tau_T$) and the particle temperature 
($T_e$) rather than specifying it a priori for spectral fitting. In addition, the model allows to constrain the size of the corona, providing 
more insight into the geometry of the region that produces the harder spectral component. Our model also includes the irradiated disk 
component, which is often missing in most works and has only been included in some \citep[e.g.][]{Vinokurov+13}. Addition of this component allows 
for truly broadband fits that include optical/UV observations which yields further constraints on the accretion disk. 

The rest of the paper is organized as follows. We present the formalism of the disk-corona model in \S\ref{sec:model} where we discuss the 
essential points and equations for the Comptonizing corona (\S\ref{sec:corona}), the disk component (\S\ref{sec:disk-spectrum}), and the 
irradiated disk contribution (\S\ref{sec:irradiated-disk}) to the total broadband spectrum. Our numerical treatment based on a one-zone 
time-dependent kinetic code is outlined in \S\ref{sec:numerical-setup}. Next, we present example model spectra, along with the corresponding 
coronal particle distribution, in \S\ref{sec:model-spectra} where we vary key model parameters to highlight their effect on the broadband spectrum. 
Constraining the mass of the central BH in the ULX Holmberg IX X-1 is the subject of \S\ref{sec:Hol-IX-X1} where we show that the X-ray observations 
favour a stellar-mass BH. We discuss the caveats and implications of our findings in \S\ref{sec:discussion} and conclude in \S\ref{sec:conclusions}.

\section{The Disk-Corona Model}\label{sec:model}

We consider an optically thick flat disk at a source distance $D$ and inclination $i$ that radiates 
a bolometric soft-photon flux $F_d = L_d\cos i/2\pi D^2$. The disk luminosity is given by 
\citep[e.g.,][]{Makishima+00}
\begin{equation}
    L_d = 4\pi (R_{\rm in}/\xi)^2\sigma_{\rm SB}(T_{\rm in}/\kappa)^4\,,
\end{equation}
where $T_{\rm in}$ is the maximum disk color temperature, $R_{\rm in}$ is the innermost disk 
radius, and $\sigma_{\rm SB}$ is the Stefan-Boltzmann constant. The correction factor $\xi = 0.41$ 
corrects for the fact that the disk attains the color temperature $T_{\rm in}$ at a slightly 
larger radius than $R_{\rm in}$ \citep[e.g.,][]{Shimura-Takahara-95}, and the factor 
$\kappa = T_{\rm in}/T_{\rm eff}\sim1.7$ relates $T_{\rm in}$ to the effective temperature 
$T_{\rm eff}$ of the disk \citep[][]{Kubota+98}. It is possible that these correction factors 
may change with the physical parameters of the system \citep{Merloni+00}. 
Here we identify $R_{\rm in}$ with the innermost 
stable circular orbit, such that $R_{\rm in}=R_{\rm ISCO}=3\alpha R_s$ where $R_s=2GM_{\rm BH}/c^2$ is 
the Schwarzschild radius, $M_{\rm BH}$ is the mass of the BH, $G$ is the gravitational constant, and $c$ 
is the speed of light. The parameter $1/6\leq\alpha\leq1$ provides a generalization for spinning (Kerr) 
BHs with $\alpha=1$ for non-spinning (Schwarzschild) BHs. It is convenient to express the disk luminosity 
using the mass of the BH, a more fundamental quantity, which gives
\begin{equation}\label{eq:L_d}
    L_d = 7.3\times10^{34}\alpha^2\fracb{\xi}{0.41}^{-2}\fracb{\kappa}{1.7}^{-4}
    \fracb{M_{\rm BH}}{10M_\odot}^2\fracb{k_BT_{\rm in}}{0.1\,{\rm keV}}^4\,{\rm erg~s}^{-1}\,.
\end{equation}
The disk luminosity is intercepted by a spherical corona with size $R_{\rm cor}$ where it is Compton 
upscattered by a hybrid (thermal + non-thermal) electron-positron coronal plasma with Thomson optical depth 
$\tau_T$. Here we make the generalization that the size of the corona is some fraction 
$\hat R_{\rm cor} = R_{\rm cor}/R_{\rm in}$ of the inner disk radius. We express the disk 
luminosity using the compactness parameter, a non-dimensional quantity that combines the luminosity and 
radius of the emission region into one \citep{Guilbert+83}, such that
\begin{eqnarray}\label{eq:ell_d}
    \ell_d &\equiv& \frac{\sigma_T}{m_ec^3}\frac{(\Omega/4\pi)L_d}{R_{\rm cor}} \\
    &=& \frac{0.2\alpha}{\hat R_{\rm cor}}\fracb{\Omega}{4\pi}\fracb{\xi}{0.41}^{-2}\fracb{\kappa}{1.7}^{-4} \nonumber
    \fracb{M_{\rm BH}}{10M_\odot}\fracb{k_BT_{\rm in}}{0.1\,{\rm keV}}^4\,,
\end{eqnarray}
where the factor of $\Omega/4\pi$ is the fraction of disk luminosity intercepted by the corona assuming that the emission is 
isotropic. The corona subtends a solid angle $\Omega = 2\pi\left(1-\sqrt{1-\hat R_{\rm cor}^2}\right)$, for $\hat R_{\rm cor}\leq1$, 
on to emitting material at a distance $R=R_{\rm in}$ from its center, which is also the center of the accreting system. In the 
two limiting cases, $\Omega = 2\pi$ when $\hat R_{\rm cor}=1$ and $\Omega = \pi\hat R_{\rm cor}^2$ when 
$\hat R_{\rm cor}\ll1$. Since most of the flux is emitted from $R\approx R_{\rm in}$ as the disk temperature declines with 
increasing $R$ (see \S\ref{sec:disk-spectrum}), we further make the crude but simpler approximation that for $\hat R_{\rm cor} > 1$, 
the solid angle is $\Omega = 4\pi$, so that the corona intercepts majority of the emitted disk luminosity. At large distances away 
from the corona, the solid angle is small and the corona intercepts only a very small fraction of emitted radiation. However, the 
contribution to the emitted flux also declines with $R$. A proper treatment of radiative transfer would require accounting for the 
radial and angular dependence of the disk emission, which cannot be done with a one-zone treatment, as adopted in this work. If 
$R_{\rm cor}<R_{\rm in}$, the spectrum should be dominated by emission arising from the disk, and therefore only show a sub-dominant 
Comptonized emission component. For the remaining discussion in this section and that in \S\ref{sec:numerical-setup} and \S\ref{sec:model-spectra}, 
we make the simplifying assumption that $R_{\rm cor}>R_{\rm in}$ and most of the disk emission is intercepted by the corona.

When the disk temperature is low, e.g. $k_BT_{\rm in}=0.1\,$keV, the disk compactness of a stellar mass BH doesn't exceed 
unity with the implication that the corona remains optically thin to scattering. However, as the mass of the BH approaches 
that of IMBHs or if the disk becomes much hotter then $\ell_d\gg1$. In this case, if a 
significant fraction of the radiation is emitted above the pair-creation threshold of 
$m_ec^2=511\,$keV, where $m_e$ is the electron rest mass, then pair-production by 
$\gamma\gamma\to e^++e^-$ becomes important and the reprocessing of the disk radiation by these 
pairs cannot be ignored. Another important consequence is the increase in optical depth of the 
pairs as $\ell_d$ becomes significantly larger than unity, which would make the corona optically 
thick if the compactness of the $e^\pm$-pairs ($\ell_e$) is at least a modest fraction of the disk compactness.

\subsection{The Comptonizing Corona}\label{sec:corona}
The corona is assumed to be loaded with $e^\pm$ pairs that have a total compactness
\begin{equation}
    \ell_e = \frac{\sigma_T}{m_ec^3}\frac{L_e}{R_{\rm cor}}\,,
\end{equation}
which is some fraction $(\ell_e/\ell_d)$ of the disk photon compactness. The injected pairs 
can be non-thermal with power-law energy distribution having number density $n_e(\gamma_e)\propto\gamma_e^{-s}$ 
for $\gamma_m\leq\gamma_e\leq\gamma_M$, where $\gamma_e=(1-\beta_e^2)^{-1/2}$ is the particle Lorentz factor bounded 
by $\gamma_m$ and $\gamma_M$, the Lorentz factors of minimal and maximal energy particles, respectively. 
In some cases, when $s$ is much larger than unity, e.g. for a very soft distribution, a delta-function is used instead. 
Magnetic fields are considered to be weak in the corona, and therefore, the pairs cool mainly by inverse-Compton 
scattering and synchrotron cooling is negligible.

Softer disk radiation with energy $E_s$ is then inverse-Compton scattered by these $e^\pm$-pairs to higher 
energies $E_c=(4/3)\gamma_e^2E_s$. The rate at which each electron is Compton cooled, thereby giving its energy to the scattered photon is 
$L_c = (4/3)\sigma_Tcp_e^2U_\gamma$, where $p_e\equiv\gamma_e\beta_e$ is the dimensionless momentum of the 
particle and $U_\gamma$ is the radiation field energy density. Then the cooling rate for the entire distribution 
of particles enclosed in a volume $(4\pi/3)R_{\rm cor}^3$ is given by 
\begin{equation}
    L_c = \frac{4}{3}\sigma_TcU_\gamma \frac{4\pi}{3}R_{\rm cor}^3\int p_e^2 n_e(\gamma_e)d\gamma_e 
    = \frac{16\pi}{9}R_{\rm cor}^2cU_\gamma\langle p_e^2\rangle\tau_T
\end{equation}
Here $\langle p_e^2\rangle$ represents the average taken over the particle distribution and $\tau_T = n_{\rm tot}\sigma_TR_{\rm cor}$ 
is the Thomson optical depth of the corona with $n_{\rm tot}$ being the total number density of particles. 
The quantity that will become useful later is the ratio of the Comptonized luminosity of the corona to the luminosity of the softer disk radiation, 
\begin{equation}\label{eq:Lc/Ld}
    \frac{L_c}{L_d} = \frac{4\xi^2\kappa^4\hat R_{\rm cor}^2c}{9\sigma_{\rm SB}}\frac{U_\gamma\langle p_e^2\rangle\tau_T}{T_{\rm in}^4}\,.
\end{equation}

\subsection{The Disk Spectrum}\label{sec:disk-spectrum}
A soft excess below $2\,$keV appears in the spectra of many ULXs \citep[e.g.,][]{Fabian-Ward-93,Kaaret+03,Miller+03} 
that is typically fit by a cool MCD spectrum \citep{Mitsuda+84} with inner disk temperature $k_BT_{\rm in}\sim0.1-0.3\,$keV. 
A similar soft excess is also seen in GBHBs, but the corresponding inner disk temperature is much higher, $k_BT_{\rm in}\sim1\,$keV.

The disk is assumed to be optically thick that radiates a multicolor blackbody spectrum  
for $R>R_{\rm in}$ with a generalized radial temperature profile, where the local disk temperature scales 
with radial distance from the BH as a power law, such that $T(R)\propto R^{-p}$ \citep{Mineshige+94}. For 
the canonical \citet{Shakura-Sunyaev-73} disk $p=3/4$. The disk spectrum for a power-law temperature profile 
can be obtained from \citep{Mitsuda+84},
\begin{equation}\label{eq:dndt-T-profile}
    \dot n_\gamma(x) = \frac{dn_\gamma}{dx\,dt} = \fracb{dn_{\gamma,\rm inj}}{dt}x^2\int_1^{\hat R_{\rm out}}\frac{\hat R}{\exp[x/\theta(\hat R)]-1}d\hat R\,,
\end{equation}
where $\dot n_\gamma(x)$ is the number density of photons per unit dimensionless energy per unit time. 
The photon energy $E$ is expressed in dimensionless units, $x=E/m_ec^2$, 
and likewise, the dimensionless temperature of the disk blackbody radiation is 
\begin{equation}
    \theta(\hat R)=\frac{k_BT(\hat R)}{m_ec^2} = \theta_{\rm in}\hat R^{-p}   
\end{equation}
where $\theta_{\rm in}$ is the temperature at the inner disk boundary at $\hat R=R/R_{\rm in}=1$. The total rate of 
photons is obtained by integrating the disk radiation up to the outer disk radius $\hat R_{\rm out} = R_{\rm out}/R_{\rm in}$.

While the softer emission (below $2\,$keV) in the majority of ULXs is well described by an MCD spectrum, 
a subset of sources show a pure blackbody-like emission with temperature 
$k_BT\approx0.05-0.2\,$ keV \citep[e.g.][]{Feng+16}. In that case, the disk blackbody spectrum with temperature 
$\theta_{\rm in}$ is given by
\begin{equation}\label{eq:dndt-BB}
    \dot n_\gamma(x) = \fracb{dn_{\gamma,\rm inj}}{dt}\frac{x^2}{\exp(x/\theta_{\rm in})-1}\,,
\end{equation}
where $dn_{\gamma,\rm inj}/dt$ is again the rate per unit volume at which soft photons are entering the corona from the disk.

\subsection{The Irradiated Disk}\label{sec:irradiated-disk}
Besides viscous heating, accretion disks are also heated by intercepting the X-ray flux from the inner disk and the 
Comptonizing corona. Following \citet{Gierlinski+09}, we construct a simple model of disk irradiation, where a fraction 
$f_{\rm out}$ of the bolometric luminosity, $L_{\rm bol}=L_d+L_c$, comprising of the inner disk and corona, is reprocessed 
by the outer disk. It is assumed here that the irradiating flux impinging at any given radius is given by 
\begin{equation}
    F_{\rm irr}=\sigma T_{\rm irr}^4(R) = \frac{L_{\rm bol}}{4\pi R^2} = \frac{L_d}{4\pi R^2}f_{\rm out}\left(1+\frac{L_c}{L_d}\right)\,,
\end{equation}
where the disk luminosity is given by Eq.~(\ref{eq:L_d}). Since the reprocessed radiation emitted by the outer disk is thermal, its 
temperature profile can now be expressed as
\begin{equation}
    \theta_{\rm irr}(\hat R) = \frac{[f_{\rm out}(1+L_c/L_d)]^{1/4}}{\kappa\xi^{1/2}}\theta_{\rm in}\hat R^{-1/2}\,,
\end{equation}
where the ratio $L_c/L_d$ is given in Eq.~(\ref{eq:Lc/Ld}) and is obtained self-consistently using the numerical treatment of 
the corona, as presented below.

\section{Numerical Treatment}\label{sec:numerical-setup}

We simulate the interactions between the photons and $e^\pm$-pairs using a one-zone kinetic code that solves the time-dependent 
coupled equations for isotropic and arbitrary distributions. The photons are initialized on a grid of dimensionless energy 
$x_{\rm min}<x<x_{\rm max}$ where the number density of photons per unit dimensionless energy at any time $t$ is given by 
$n_\gamma(x,t)=\partial n_\gamma/\partial x$. Likewise, the $e^\pm$-pairs are initialized with a grid in dimensionless 
momentum with $p_{\rm min}<p_e<p_{\rm max}$ where their number density per unit dimensionless momentum is given by 
$n_\pm(p_e,t)=\partial n_\pm/\partial p_e$. Since the code has no spatial information, the simulated region, which represents 
the corona, is assumed to be spherical with radial extent $R_{\rm cor}$. The code solves the following coupled equations,
\begin{align}
    \dot n_\gamma(x,t) & = \dot n_{\gamma,\rm inj}(x,t) + \dot n_{\gamma,\rm cs}(x,t) + \dot n_{\gamma,\rm pp}(x,t) 
    + \dot n_{\gamma,\rm pa}(x,t) \\
    &+ \dot n_{\gamma,\rm esc}(x,t) \nonumber \\
    \dot n_\pm(p_e,t) & = \dot n_{\pm,\rm inj}(p_e,t) + \dot n_{\pm,\rm cs}(p_e,t) + \dot n_{\pm,\rm pp}(p_e,t) 
    + \dot n_{\pm,\rm pa}(p_e,t) \\
    &+ \dot n_{\pm,\rm coul}(p_e,t) \nonumber \,,
\end{align}
where the different terms represent interaction rates per unit volume per unit dimensionless energy (momentum) for photons 
(particles) \citep[see, e.g.,][for the different rates and their numerical implementation]{Belmont+08,Belmont-09,Vurm-Poutanen-09,Gill-Thompson-14}. 
Both photons and particle equations include terms that represent the rates for injection (inj), Compton scattering 
(cs; $\gamma + e^\pm \to \gamma + e^\pm$), pair-production (pp; $\gamma + \gamma \to e^+ + e^-$), and pair-annihilation 
(pa; $e^+ + e^- \to 2\gamma$). Additional terms include an escape (esc) term for photons (particles don't escape the corona) and a 
Coulomb (coul; $e^+ + e^- \to e^+ + e^-$) interaction term for the particles.

The multi-color thermal emission emerging from the disk is intercepted by the corona where it is Comptonized by the $e^\pm$-pairs. The rate 
of injection of these thermal photons is set by the compactness, which in turn is set by the black hole mass $M_{\rm BH}$ and the fractional 
size of the corona $\hat R_{\rm cor}$ in Eq.~(\ref{eq:ell_d}), 
to give
\begin{equation}
    \ell_d = \frac{4\pi\sigma_T}{3c}R_{\rm cor}^2\int\dot n_{\gamma,\rm inj}(x)x\,dx\,,
\end{equation}
where we have assumed a constant rate of injection. The above equation is then used to determine the normalization $dn_{\gamma,\rm inj}/dt$ in 
Eq.~(\ref{eq:dndt-T-profile} \& \ref{eq:dndt-BB}) for a given $\theta_{\rm in}$, $p$, and $\hat R_{\rm out}$.

\begin{figure*}
    \centering
    \includegraphics[width=0.48\textwidth]{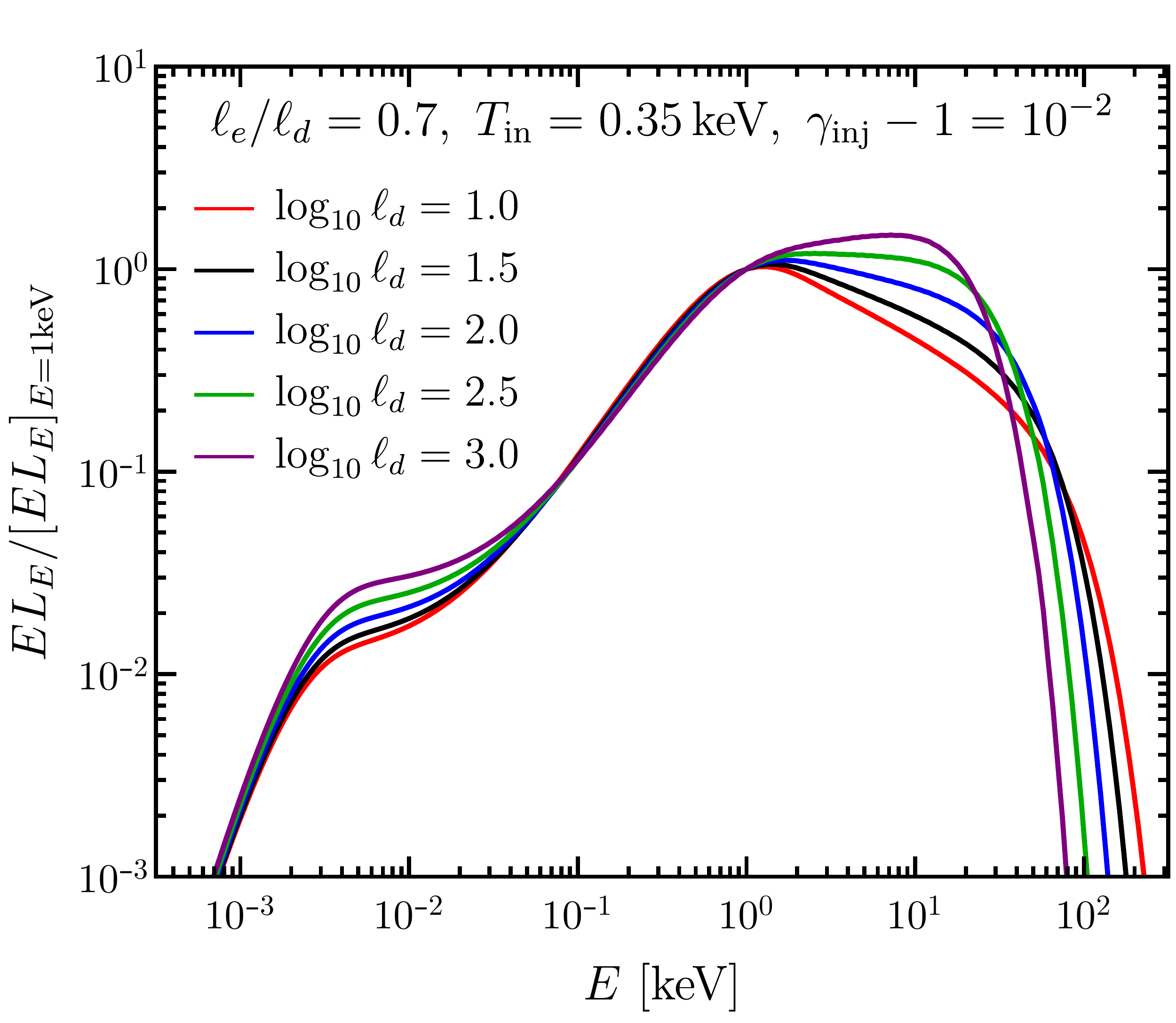}\quad\quad
    \includegraphics[width=0.48\textwidth]{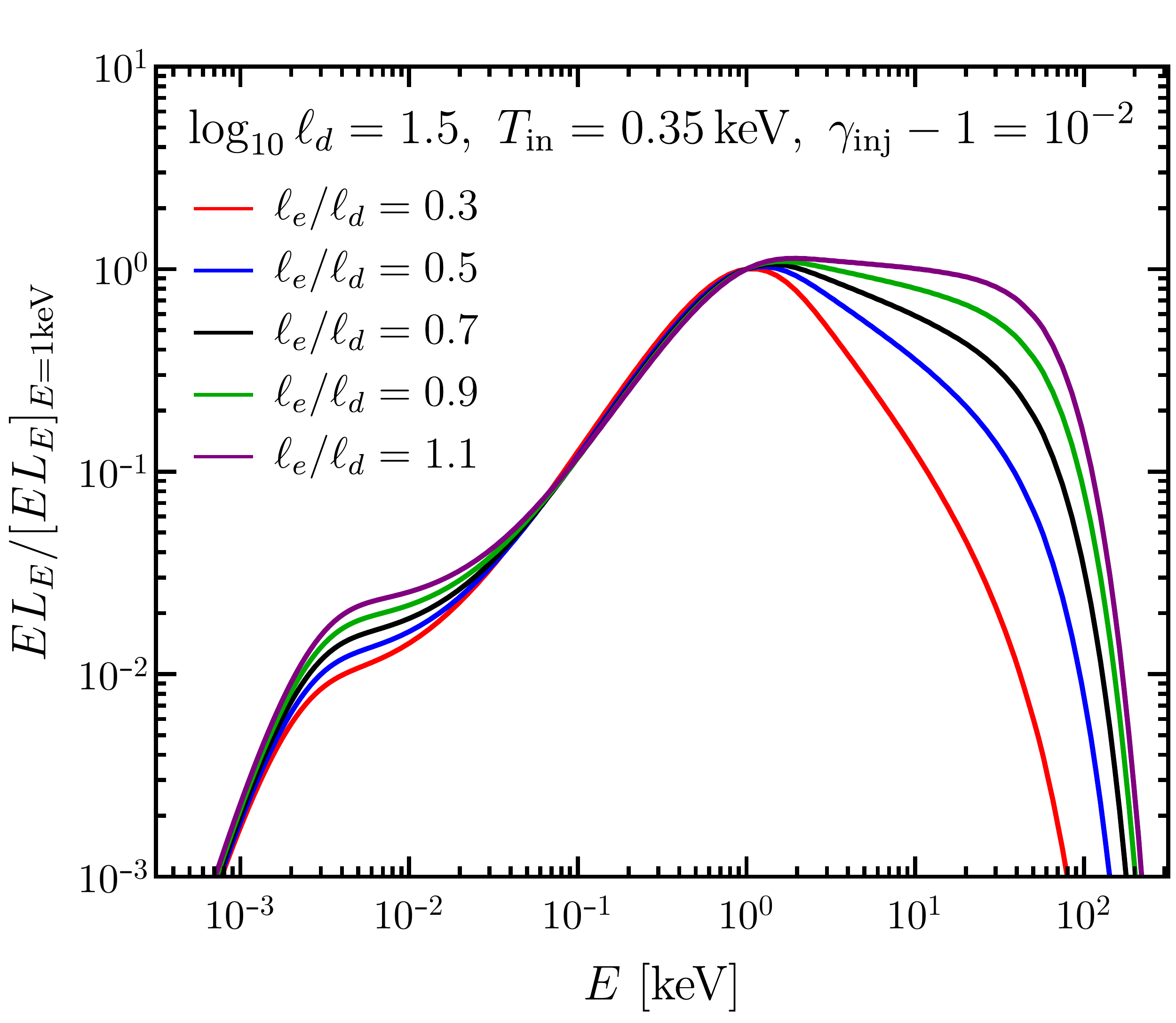}
    \includegraphics[width=0.48\textwidth]{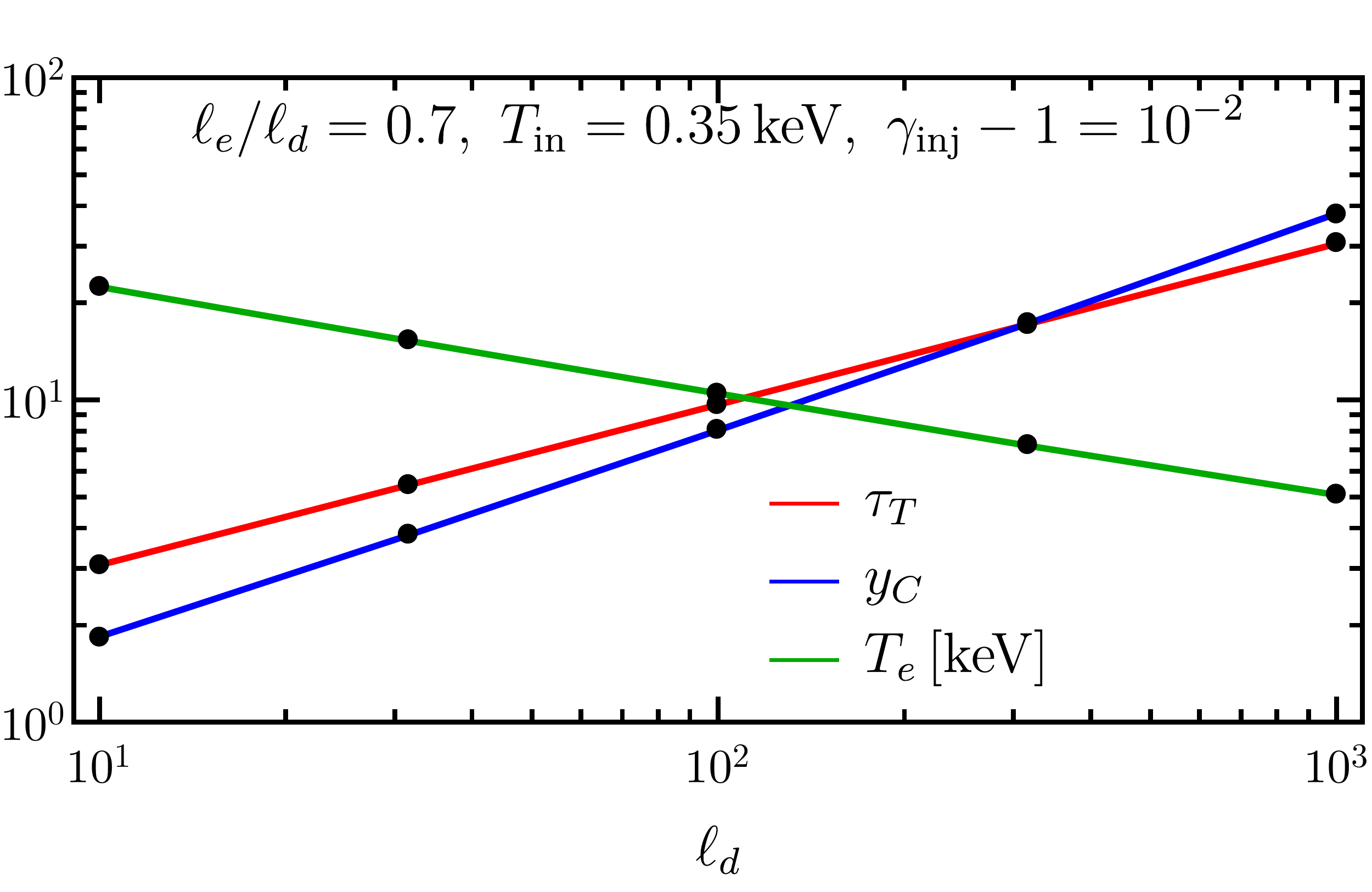}\quad\quad
    \includegraphics[width=0.48\textwidth]{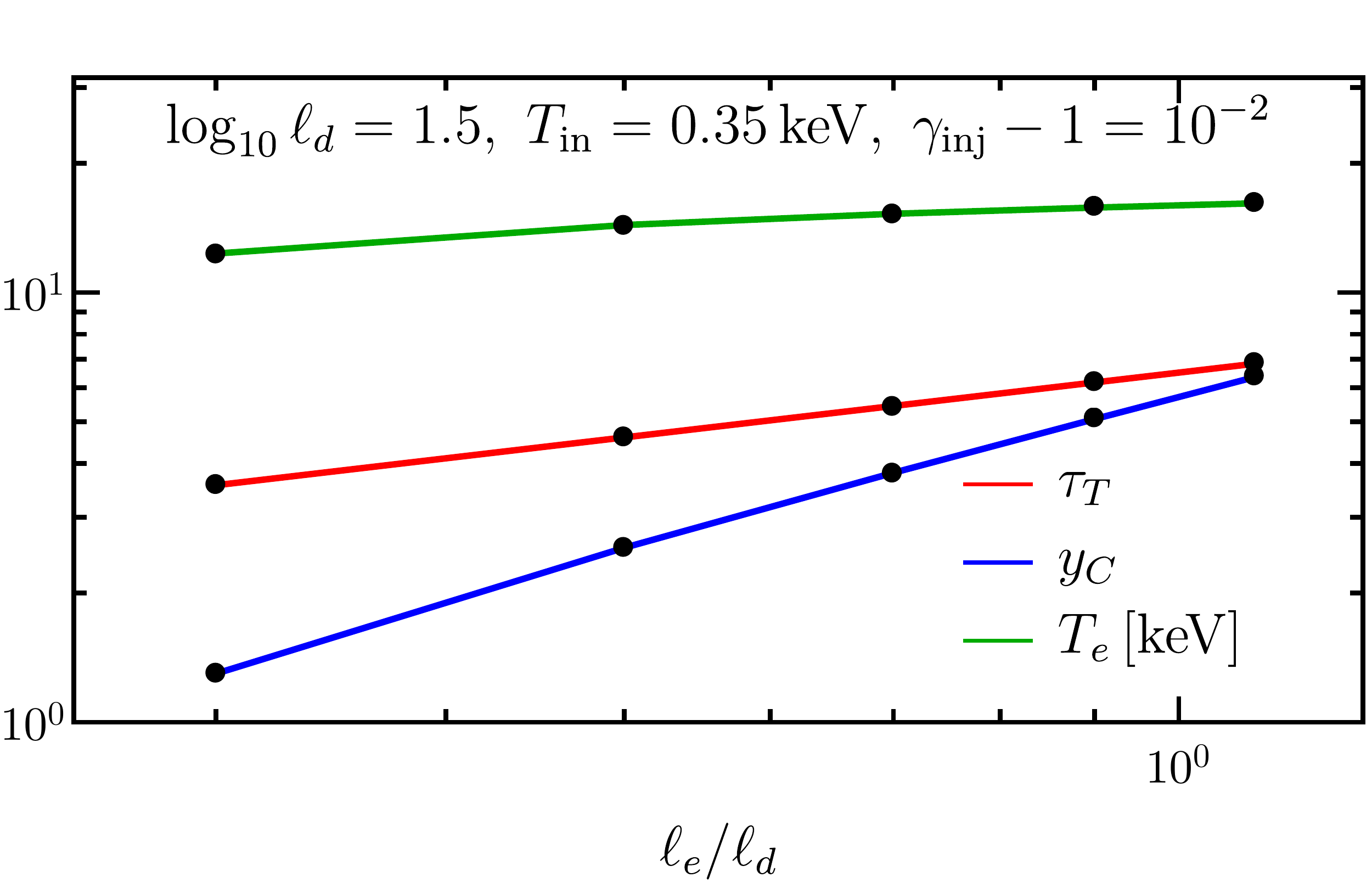}
    \caption{Steady-state model spectra for different disk compactness $\ell_d$ (top-left; obtained by changing $\hat M_{\rm BH}$), 
    and different compactness ratios $\ell_e/\ell_d$ (top-right). The non-relativistic $e^\pm$-pairs are injected as a delta-function with 
    injection LF $\gamma_{\rm inj}$ and the disk photons have an MCD spectrum with $p=0.75$. For clarity all the spectra are normalized 
    by their luminosity at $E=1\,$keV and the pair-annihilation line at $E=511\,$keV has been removed. Other fiducial model parameters assumed 
    here are: $\xi=0.41$, $\kappa=1.7$, $f_{\rm out}=10^{-2}$, $\hat R_{\rm out}=5\times10^4$, $\hat R_{\rm cor}=10$. 
    }
    \label{fig:spec-diff-ld-leld}
\end{figure*}

The injection of $e^\pm$-pairs into the corona is determined in a similar manner, where the rate of injection (also assumed to be constant here) 
is controlled by the pair compactness,
\begin{equation}
    \frac{\ell_e}{2} = \frac{4\pi\sigma_T}{3c}R_{\rm cor}^2\int\dot n_{\pm,\rm inj}(p_e)\gamma_e\,dp_e\,.
\end{equation}
Here the compactness also includes the rest mass of particles, and not just the kinetic energy, and the factor 
of (1/2) multiplying $\ell_e$ means that only half of the total compactness goes into each species (electron and positron) of particles. 
When considering a power-law energy distribution of pairs, such as 
$\dot n_\pm(p_e) = (p_e/\gamma_e)\dot n_\pm(\gamma_e) = (p_e/\gamma_e) (dn_{\pm,\rm inj}/dt)\gamma_e^{-s}$ 
for $\gamma_{\rm min}\leq\gamma_e<\leq\gamma_{\rm max}$, the normalization $dn_{\pm,\rm inj}/dt$ can be obtained from the above equation that yields
\begin{equation}
    \frac{dn_{\pm,\rm inj}}{dt} = \frac{3c}{8\pi\sigma_T}\frac{\ell_e}{R_{\rm cor}^2}
    \left[\frac{s-2}{\gamma_{\rm min}^{2-s}-\gamma_{\rm max}^{2-s}}\right]~~(s>2)\,.
\end{equation}
For a delta-function injection of pairs with distribution $\dot n_\pm(p_e)=(p_e/\gamma_e)(dn_\pm/dt)\delta(\gamma_e-\gamma_0)$, the 
rate of injection is given by
\begin{equation}\label{eq:delta-func-injection-rate}
    \frac{dn_{\pm,\rm inj}}{dt} = \frac{3c}{8\pi\sigma_T}\frac{\ell_e}{R_{\rm cor}^2\gamma_0}\,,
\end{equation}
where $\gamma_0$ is the energy per unit rest mass energy per particle of the injected pairs.

The Comptonized radiation continuously escapes the corona with rate 
\begin{equation}
    \dot n_{\gamma,\rm esc}(x,t)=-\kappa(x,t)\frac{n_\gamma(x,t)}{t_{\rm dyn}}\,,
\end{equation}
where we use the standard `leaky-box' approximation under which the radiation leaks out with an escape probability $\kappa(x,t)$ over 
the dynamical time $t_{\rm dyn}=R_{\rm cor}/c$. For a spherical region $\kappa(x,t)$ is given by Eq.~(21) of \citet{Lightman-Zdziarski-87}. 

We compare the results of our numerical model with that from CompPS \citep{Poutanen-Svensson-96} for a simple Comptonization problem in 
Appendix~\ref{sec:Model-Comparison} and find good agreement.

\begin{figure*}
    \centering
    \includegraphics[width=0.48\textwidth]{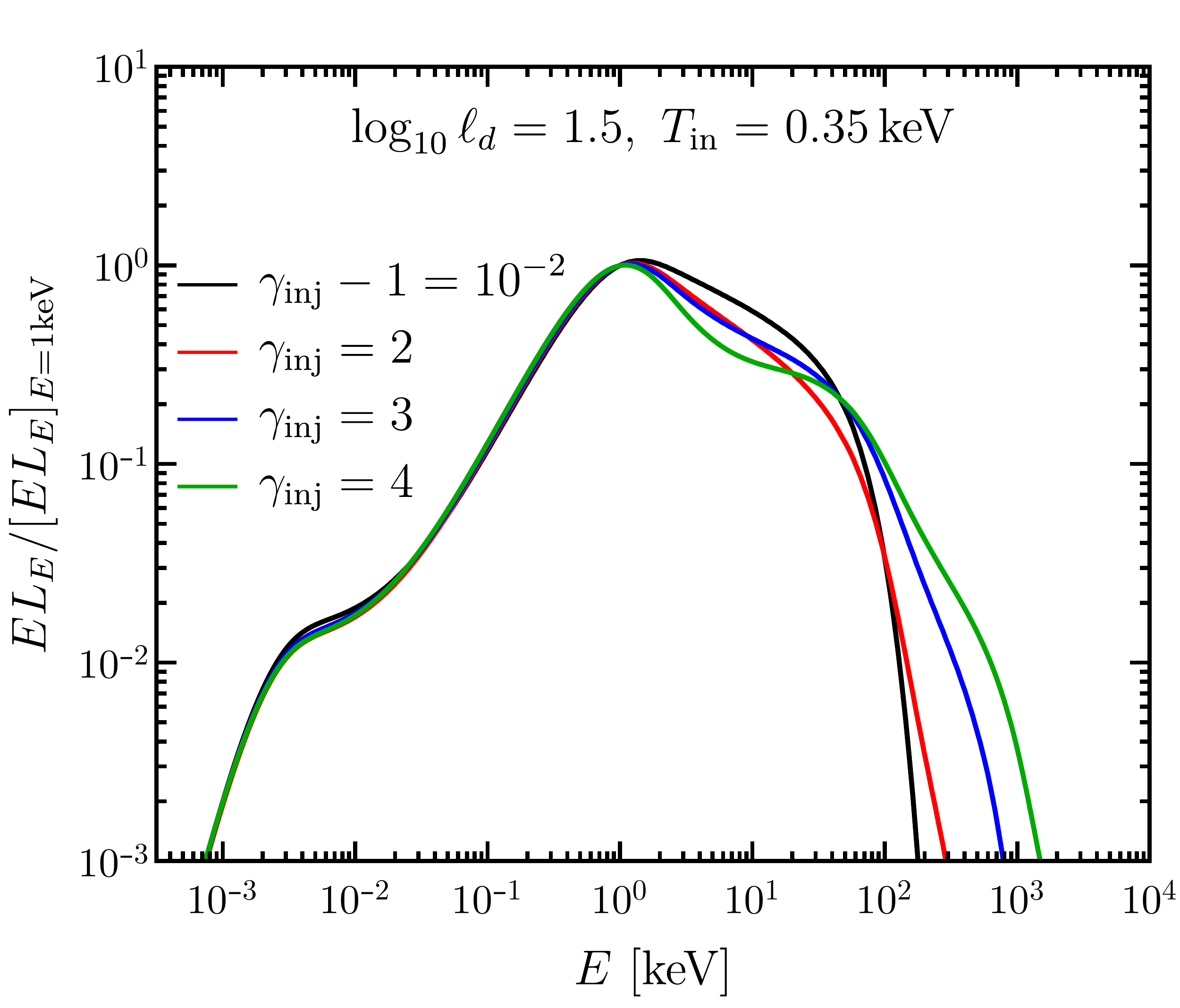}\quad\quad
    \includegraphics[width=0.48\textwidth]{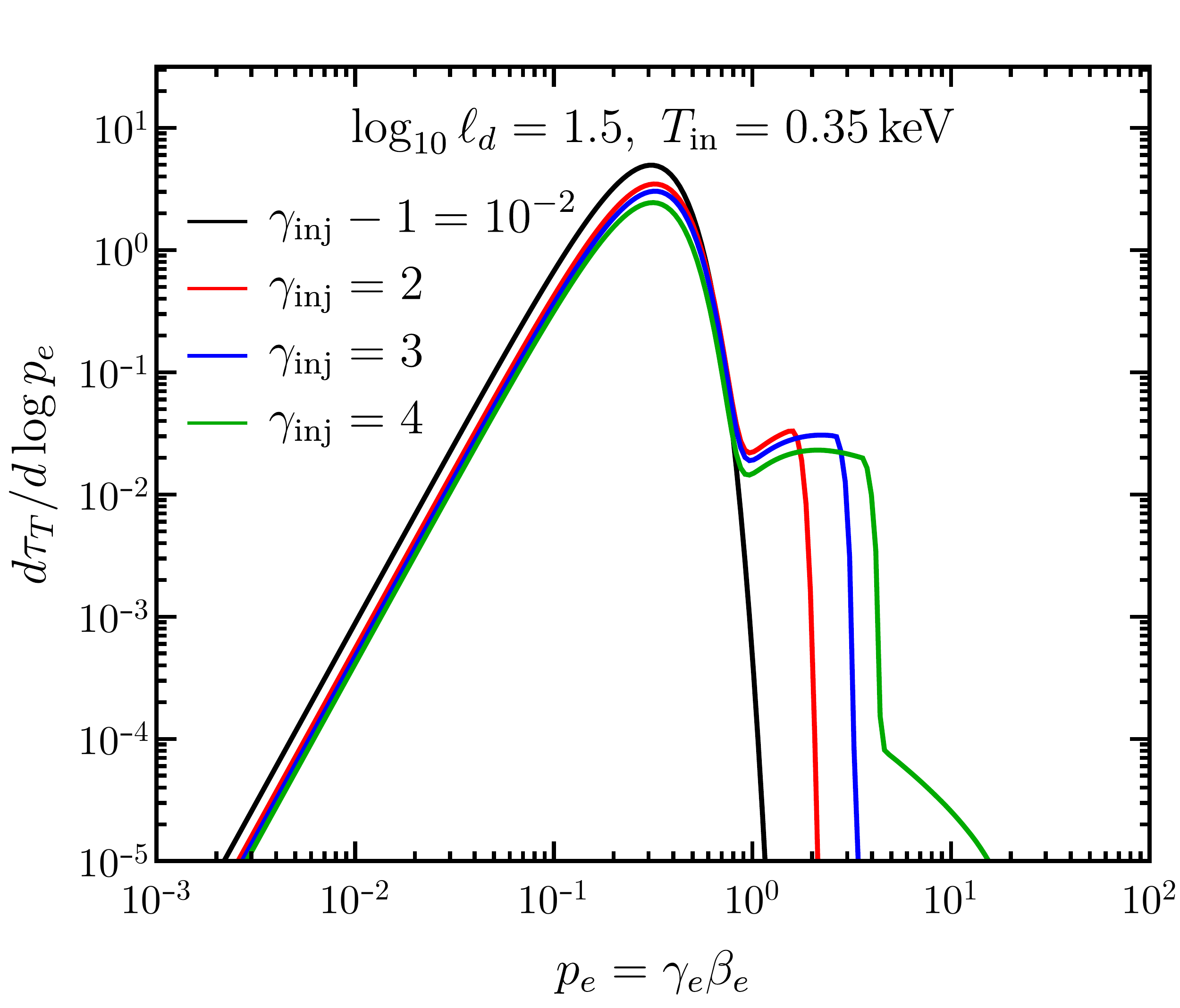}
    \caption{(Left) Steady state model spectra with different $\gamma_{\rm inj}$ for a $\delta$-function particle injection. 
    (Right) The corresponding steady state particle distribution. All other parameters are the same as in Fig.~\ref{fig:spec-diff-ld-leld}.}
    \label{fig:diff-ginj}
\end{figure*}

\begin{figure*}
    \centering
    \includegraphics[width=0.48\textwidth]{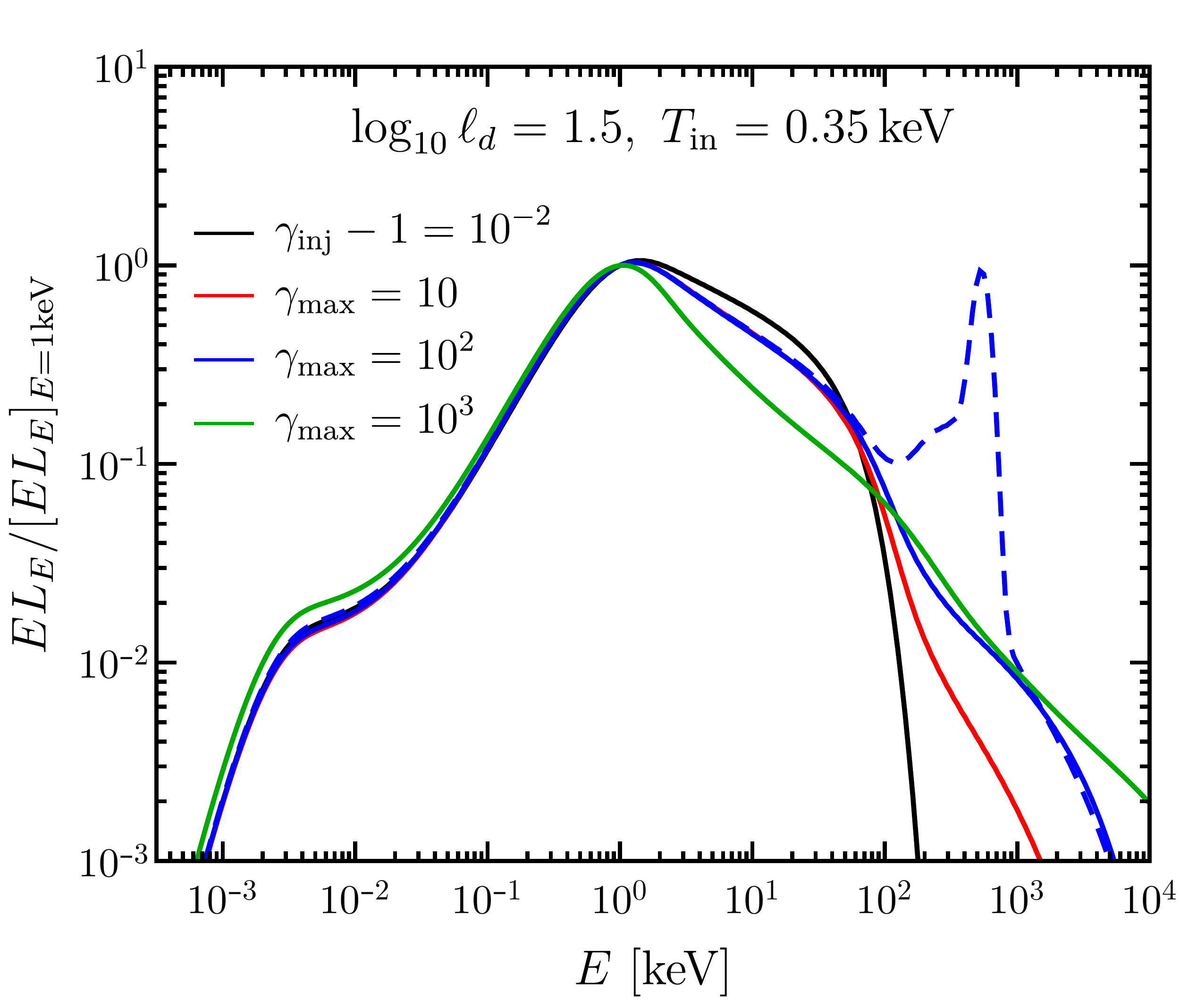}\quad\quad
    \includegraphics[width=0.48\textwidth]{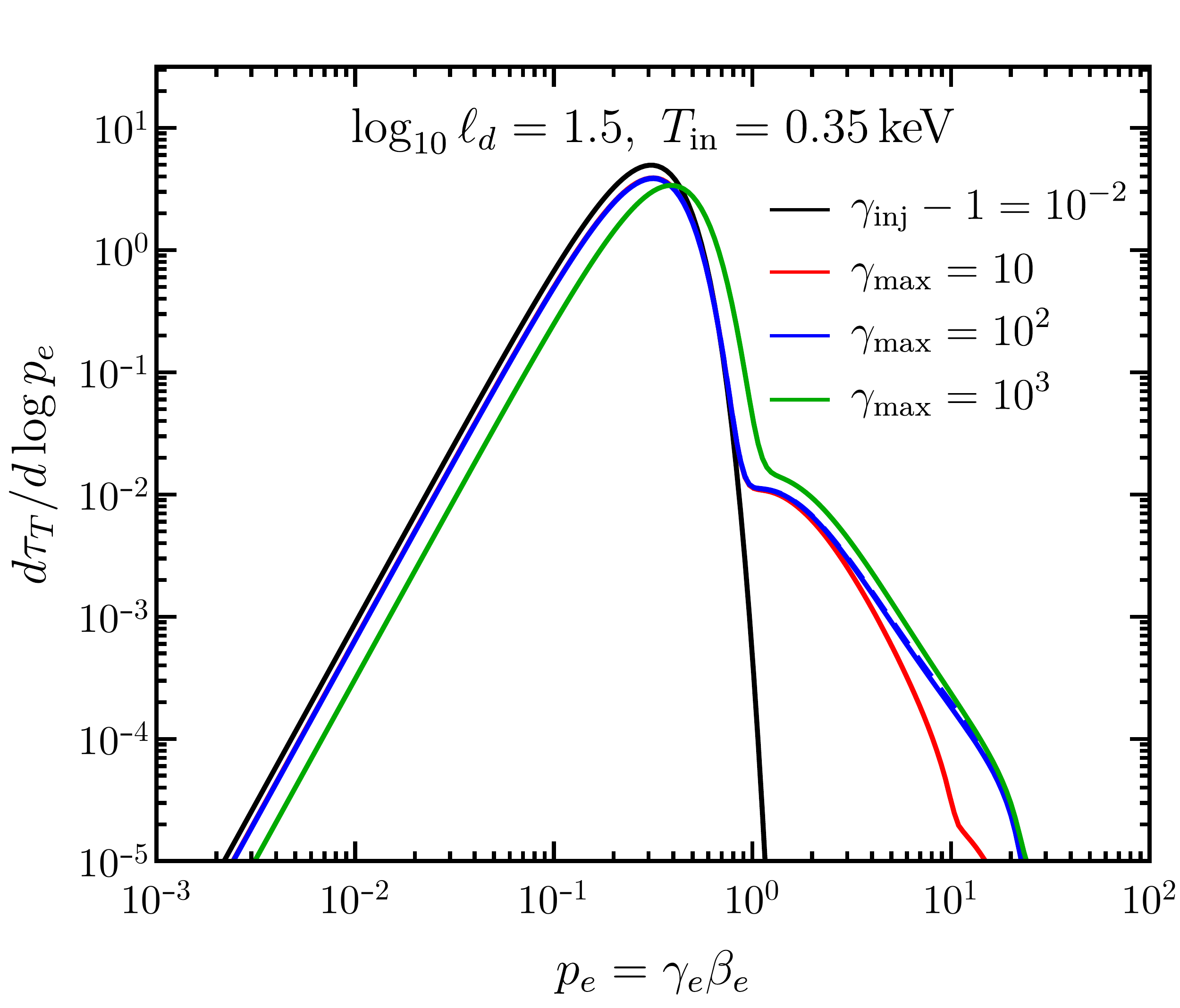}
    \caption{(Left) Steady state model spectra with different $\gamma_{\rm max}$ for a power-law particle distribution injection 
    with $\gamma_{\rm min}=1$ and power-law index $s=3.5$. The dashed curve shows the spectrum with contribution from the $e^\pm$-pair 
    annihilation line included. 
    (Right) The corresponding steady state particle distribution. All other parameters are the same as in Fig.~\ref{fig:spec-diff-ld-leld}.}
    \label{fig:diff-gmax}
\end{figure*}

\section{Model Spectra}\label{sec:model-spectra}
Our model has in total $8-11$ parameters that control different features of the emergent spectrum: 
\begin{enumerate}
    \item $\hat M_{\rm BH}\equiv\alpha M_{\rm BH}/M_\odot$: Normalized BH mass
    \item $T_{\rm in}$: Disk temperature at the inner disk radius $R_{\rm in}$
    \item $\hat R_{\rm cor}$: Fractional size of the corona in terms of $R_{\rm in}$
    \item $\ell_e/\ell_d$: the ratio of the particle ($e^\pm$-pairs) compactness to that of the disk radiation, where the latter, 
    $\ell_d\propto \hat M_{\rm BH}T_{\rm in}^4/\hat R_{\rm cor}$, depends on the combination of other three parameters
    \item $p$: Radial power-law index of the disk temperature
    \item $f_{\rm out}$: Fraction of the comptonized disk radiation intercepted by the outer disk
    \item $\hat R_{\rm out}$: Outer disk radius in terms of $R_{\rm in}$
    \item $\gamma_{\rm inj}$: Lorentz factor of the $e^\pm$-pairs injected into the corona when using a $\delta$-function distribution
    \item $\gamma_M$: Maximum Lorentz factor of the $e^\pm$-pairs when using a power-law distribution with $\gamma_m=1$
    \item $s$: Power-law index of the injected $e^\pm$-pairs
    \item $N_H$: Column density of the ISM; in some cases it represents only the extra-Galactic column density
\end{enumerate}
The number of parameters vary depending on whether the particle distribution is a power law or delta-function. 
Below we show model spectra by varying those model parameters whose effect is non-trivial.

The left column of Fig.~\ref{fig:spec-diff-ld-leld} shows the effect of varying $\ell_d$ which can be obtained by varying the three 
fundamental model parameters, namely $\hat M_{\rm BH}$, $T_{\rm in}$, $\hat R_{\rm cor}$, as can be seen from Eq.~(\ref{eq:ell_d}); changes in 
$T_{\rm in}$ would also shift the spectral peak of the disk emission. The $\nu F_\nu$ thermal peak appears approximately at 
$E_{\rm pk}\simeq2.82k_BT_{\rm in}$, and 
not exactly at this energy due to some broadening of the peak caused by Comptonization. As the disk radiation compactness $\ell_d$ is increased, 
the Comptonized spectrum above the thermal peak becomes harder. Since the Thomson optical depth of pairs in the corona is $\tau_T>1$, 
the particle distribution thermalizes at $T_e>T_{\rm in}$, where $T_e$ is the temperature of the pairs. $T_e$ is sensitive to $\tau_T$ 
and a larger $\tau_T$, which for a fixed $\hat R_{\rm cor}$ would mean larger number of particles, would lead to a smaller mean energy 
per particle and therefore lower $T_e$. 
The importance of Comptonization is determined by the Compton-$y$ parameter, $y_C=(4/3)\langle p_e^2\rangle\max(\tau_T,\tau_T^2)$, where 
$\langle p_e^2\rangle = 3\theta_e$ for a non-relativistic Maxwellian, with particle non-dimensional temperature $\theta_e \equiv k_BT_e/m_ec^2<1$. 
We show the trend of $\tau_T$, $y_C$, and $T_e$ with disk emission compactness $\ell_d$ in the bottom row of Fig.~\ref{fig:spec-diff-ld-leld}. 
When $\tau_T\gtrsim1$ and $y_C\gtrsim1$, Comptonization proceeds in the unsaturated regime \citep[e.g.,][]{Rybicki-Lightman-86} and 
the formation of a Wien peak at $E=3k_BT_e$ is delayed until $y_C\gg1$ (saturated Comptonization). In the present case, 
the spectrum shows a power-law above the thermal peak energy and an exponential cutoff (a Wien tail) at $E\approx3k_BT_e$. Therefore, 
as $\tau_T$ increases due to increasing $\ell_d$, the particle temperature declines and the exponential cutoff accordingly moves to 
lower energies while the spectrum becomes harder due to the rising $y_C$.

The effect of varying the compactness ratio $\ell_e/\ell_d$ is demonstrated in the right column of Fig.~\ref{fig:spec-diff-ld-leld} for 
a fixed $\ell_d$ with the same rest of the parameters. The spectral trend is similar to that shown in the top-left panel, however, 
with one important difference. In this case, as can be seen from the bottom-right panel, the temperature remains approximately steady 
since the change in $\tau_T$ is rather modest. Consequently, the exponential break remains at the same energy for different values of 
$\ell_e/\ell_d$. The spectrum still hardens due to the increasing particle compactness $\ell_e$ which yields larger $y_C$.

In both cases, the spectrum remains below the $\gamma\gamma$ pair-creation threshold of $E_{\gamma\gamma}\approx m_ec^2$, and therefore, 
apart from the injected pairs there is no additional pair creation. Since the pairs remain non-relativistic, with $\theta_e<1$, they 
do annihilate. For a warm plasma, the pair-annihilation cross-section can be approximated as 
$\langle\sigma_{\rm ann}\vert\vec v_+-\vec v_-\vert\rangle\simeq (3/8)\sigma_Tc$. This yields the rate of annihilation in terms of the 
rate of decline in the total optical depth
\begin{equation}
    \frac{d\tau_T}{dt}\vert_{\rm ann} \simeq \frac{3}{16}\frac{\tau_T^2}{(R_{\rm cor}/c)}\,,
\end{equation}
which increases quadratically with increasing optical depth. By comparing the rate of injection from Eq.~(\ref{eq:delta-func-injection-rate}), 
e.g. for a delta-function injection case, $d\tau_{T,\rm inj}/dt = 2\sigma_TR_{\rm cor}dn_{\pm,\rm inj}/dt$, we can derive the condition 
for steady-state at which point the injection and annihilation rates are equal,
\begin{eqnarray}
    &&\tau_{T,\rm steady~state} = \sqrt{\frac{4}{\pi}\frac{\ell_e}{\gamma_{\rm inj}}} \\
    && \approx 0.6\left[\frac{(\ell_e/\ell_d)}{\gamma_{\rm inj}\hat R_{\rm cor}}\right]^{1/2}\fracb{\xi}{0.41}^{-1}\fracb{\kappa}{1.7}^{-2}
    \fracb{\hat M_{\rm BH}}{10}^{1/2}\fracb{k_BT_{\rm in}}{0.1\,{\rm keV}}^2\,. \nonumber
\end{eqnarray}
Another common feature in both scenarios is the rising normalization of the irradiated disk component. This occurs due to the fact that 
$L_c/L_d$ scales positively with both $\ell_d$ and $\ell_e/\ell_d$, which changes the normalization of the irradiated disk temperature $\theta_{\rm irr}$.

The above equation for $\tau_{T,\rm steady~state}$ cannot be strictly used to constrain the coronal properties in other accreting systems, 
particularly in active galactic nuceli (AGN), since it neglects the creation of $e^\pm$ pairs via $\gamma\gamma$-annihilation, which would necessarily 
occur in AGN coronae. However, order of magnitude 
estimates can still be obtained. For example, the hot coronae of AGNs with electron temperatures $k_BT_e\sim(10-500)\,$keV have optical depths 
of $\tau_T\sim0.1-4$ \citep[e.g.,][]{Fabian+17,Tortosa+18}. The soft photon (blackbody disk emission) spectrum in these systems have characteristic 
temperatures of $k_BT_{\rm in}\sim5\,$eV \citep[e.g.,][]{Shang+05} and the system is typically not so much photon starved with $\ell_e/\ell_d\sim5$. 
Then for $\hat M_{\rm BH}=10^{8.5}$, $\gamma_{\rm inj}=10^3$, and $\hat R_{\rm cor}=1$, we get $\tau_{T,\rm steady~state}=0.6$, which is consistent 
with order unity coronal optical depths inferred in AGNs.

The effect of increasing the Lorentz factor $\gamma_{\rm inj}$ of the injected pairs for a $\delta$-function distribution is shown in 
Fig.~\ref{fig:diff-ginj}. The left panel of the figure compares the injection of relativistic monoenergetic particles with the fiducial 
case where the injected particles are non-relativistic with $\gamma_{\rm inj}-1=10^{-2}$. As $\gamma_{\rm inj}$ is increased the 
spectrum above the thermal peak becomes softer as the inverse-Compton scattered peak, at $E_C=(4/3)\gamma_{\rm inj}^2E_{\rm th}$, moves 
to larger energies. A secondary inverse-Compton scattered peak is also visible at higher energies. The right panel of the figure shows 
the corresponding steady-state particle distribution which remains predominantly thermal due to the large compactness of the thermal 
disk emission and $\tau_T\gtrsim1$.

Next, we inject a power-law distribution of particles into the corona with $s=3.5$ and varying $\gamma_{\rm max}$ while keeping $\gamma_{\rm min}=1$. 
The steady-state spectra and particle distributions are shown in Fig.~\ref{fig:diff-gmax}. 
Even here the spectrum above the thermal peak is softer than the fiducial case (black curve), but it extends to increasingly larger energies 
as $\gamma_{\rm max}$ is increased. The high-energy tail is also shaped by $\gamma\gamma$-annihilation as the spectral luminosity 
increases above the pair-creation threshold of $\sim m_ec^2$. In all the spectra shown here, we have removed the contribution of the 
pair annihilation line that appears at $E=511\,$keV. How the spectrum would change if it was included is demonstrated for one of the cases 
(blue dashed curve). Since there is no particle escape from the corona, $e^\pm$-pair annihilation acts as a way to remove particles so that 
a steady state can be obtained. The particle distribution now shows a distinct high-energy tail that constitutes the steady-state distribution of 
the cooled pairs. This results in a `hybrid' distribution \citep{Coppi-00} which has both thermal and non-thermal components.

\begin{figure}
    \centering
    \includegraphics[width=0.48\textwidth]{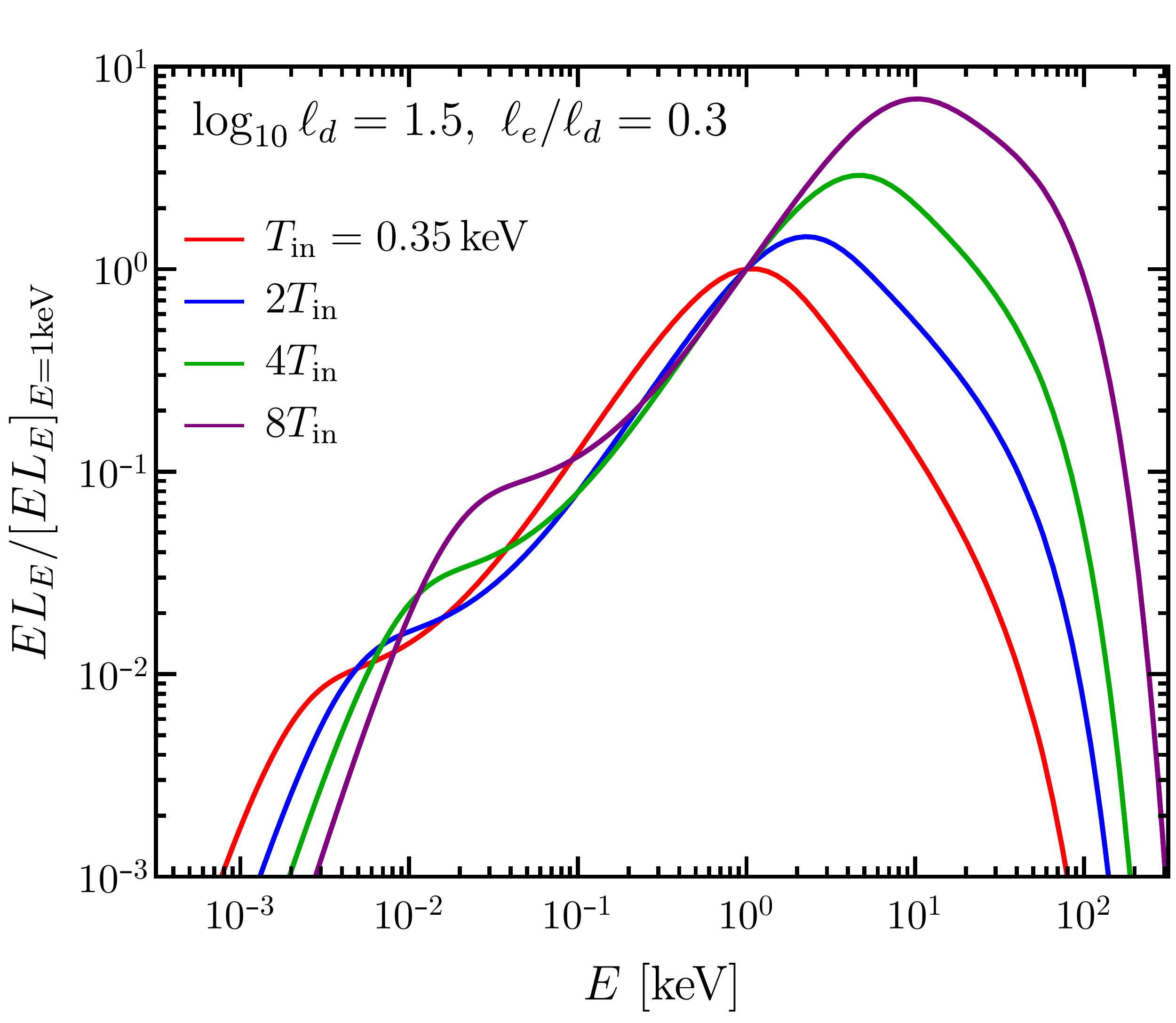}
    \caption{Steady-state model spectra with different inner disk temperature $T_{\rm in}$, and with $\gamma_{\rm inj}=1+(3/2)\theta_{\rm inj}$ where 
    $\theta_{\rm inj}=10k_BT_{\rm in}/m_ec^2$. The rest of the parameters, except $\hat M_{\rm BH}$ and/or $\hat R_{\rm cor}$, are the same as 
    in Fig.~\ref{fig:spec-diff-ld-leld}. $\hat M_{\rm BH}$ and/or $\hat R_{\rm cor}$ were changed accordingly in order to keep a fixed $\ell_d$.}
    \label{fig:diff-Tin}
\end{figure}

\begin{figure}
    \centering
    \includegraphics[width=0.48\textwidth]{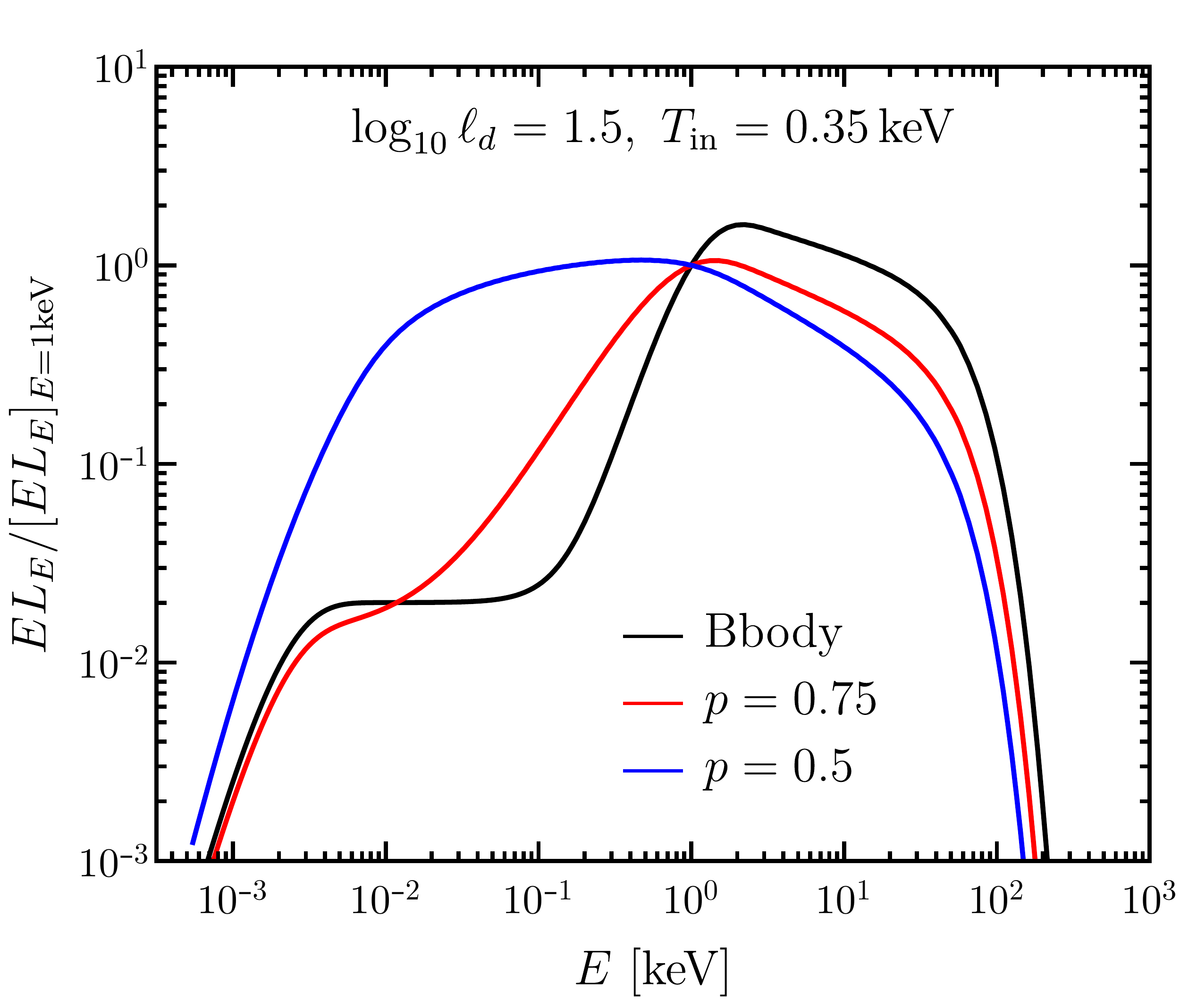}
    \caption{Comparison of steady-state spectra with different temperature profile for the disk, $\theta(\hat R)\propto\hat R^{-p}$. 
    The multicolor disk spectra are compared with the case of blackbody emission coming from the inner-disk radius. 
    The rest of the parameters are the same as in Fig.~\ref{fig:spec-diff-ld-leld}.}
    \label{fig:diff-pBB}
\end{figure}

In Fig.~\ref{fig:diff-Tin}, we show that as long as the disk and coronal plasma compactness ($\ell_d$ and $\ell_e$) are kept fixed, the steady 
state spectral shape remains approximately constant, apart from a translation of the spectrum to higher energies as well as towards higher luminosities, both 
owing to the rising $T_{\rm in}$. As a result, many of the example model spectra shown for $T_{\rm in}=0.35\,$keV can be approximately obtained for higher 
$T_{\rm in}$ with an appropriate shift in the spectral peak energy and normalization. In all the cases shown in Fig.~\ref{fig:diff-Tin} $\tau_T$ and the 
Compton-$y$ parameters are very similar.

The effect of changing the temperature profile of the disk, $\theta(\hat R)\propto\hat R^{-p}$, with different power-law indices is shown in 
Fig.~\ref{fig:diff-pBB}. A harder temperature profile causes the flux to shift to lower energies thereby softening the spectrum below the 
thermal peak. 

\section{Holmberg IX X-1: BH Mass Estimates}\label{sec:Hol-IX-X1}
Holmberg IX X-1 is one of the most widely studied bright ULXs in the last two decades 
\citep[e.g.,][]{LaParola+01,Wang-02,Miller+04,Dewangan+06,Tsunoda+06,Gladstone+09,Kaaret-Feng-09,Kong+10,Vierdayanti+10,Walton+13,Walton+14,Luangtip+16,Walton+17} 
owing to its extreme persistent luminosity of $L_X>10^{40}\,{\rm erg\,s}^{-1}$. It is associated with the dwarf galaxy Holmberg IX, 
a satellite of the spiral galaxy M81, at a distance of $D\simeq3.55\,$ Mpc. Initial spectral modeling of Holmberg IX X-1, that utilized the 
XMM-\textit{Newton} data, revealed a soft, with $k_BT_{\rm in}\sim(0.17-0.29)\,$keV, disk component plus power-law continuum that completely dominates the 
spectrum at $E\gtrsim2\,$keV \citep{Miller+04}. Based on this it was argued that if the ULXs follow the scaling relation, $T_{\rm in}\propto M_{\rm BH}^{-1/2}$, 
as is true for the GBHBs, then Holmberg IX X-1 must harbor a BH with mass $M_{\rm BH}\sim{\rm few}\times10^2M_\odot$ if it is accreting at or near $L_{\rm Edd}$. 
The IMBH interpretation precludes the presence of spectral curvature in the hard spectrum, which was later discovered in the $\sim2-10\,$keV energy band 
the origin of which was found to be optically-thick ($\tau_T\sim5-30$) coronae \citep{Stobbart+06,Dewangan+06,Gladstone+09,Walton+13}. A disk plus thermal Comptonization 
model, e.g. compTT \citep{Titarchuk-94} and EQPAIR \citep{Coppi-00}, was used to successfully describe the curved spectrum of Holmberg IX X-1 
where the temperature of the Comptonizing corona was found to be $kT_e\sim(2-3)\,$ keV. When \textit{NuSTAR} data became available for this source, 
the hard excess above $10\,$keV was modeled using the SIMPL Comptonization or CUTTOFFPL models \citep[][]{Walton+14,Luangtip+16,Walton+17,Walton+18}.

In the following, we first describe the observations and data reduction procedures for the three different spectra of Holmberg IX X-1 dubbed 
LOW, MEDIUM, and HIGH based on their relative X-ray fluxes. After that we present the spectral analysis and fit results.

\subsection{Observations and Data Reduction}
As part of our study of the spectral variability of Holmberg IX X-1, we have analyzed (following \citealt{Walton+14}) coordinated broadband X-ray 
\textit{NuSTAR}, \textit{XMM-Newton}, and \textit{Suzaku} observations for 3 epochs spanning the period 2012-2015. For the analysis, we deployed 
the most up-to-date calibration files for all the instruments. We note that the Galactic absorption component was removed from all low-energy data 
by using the responses for each detector. Below, we outline our data reduction procedures for these observations.

\subsubsection{XMM–Newton}
The \textit{XMM–Newton} data for the first two epochs (i.e., high flux and medium flux; ObsIDs 0693851701 and 0693850901 respectively) were reduced, and source and background spectra, as well as, their redistribution matrices and auxiliary response files (RMF and ARF) were created using the remote interface SAS analysis tool RISA for EPIC-pn and EPIC-MOS detectors. We deployed the pattern 0-4 for pn, and 0-12. for the MOS. To avoid pixels close to CCD boundaries and dead columns FLAG was set to 0. The source spectra were extracted using a circular radius of $\sim$30$\arcsec$ and care was taken to avoid chip gaps. In all cases the background was extracted using a larger area of the same CCD frame but in a region free from point sources. Finally, the spectra were rebinned using the grppha task in HEASOFT to have a minimum of 50 counts per bin.

\subsubsection{Suzaku}
We followed the procedure outlined in the \textit{Suzaku} data-analysis guide\footnote{http://heasarc.gsfc.nasa.gov/docs/suzaku/analysis/} to reduce 
the epoch 4 data by using $\heasoft~v2.26.2$ for the front illuminated XIS0 and XIS3 units over the $0.6-10.0\,$keV range and XIS1 unit over the 
$0.7-9.0\,$keV range respectively. Following \citealt{Walton+14}, we excluded the $1.7-2.1\,$keV range in all three detectors so as to avoid known calibration issues around the instrumental edges. The source spectra were extracted using a circular radius of $\sim$70$\arcsec$ and a larger source-free region was used to extract the background. All spectra were extracted from cleaned event files using XSELECT. All XIS spectra were rebinned using the grppha task in HEASOFT to have a minimum of 50 counts per bin.  

\subsubsection{NuSTAR}
Corresponding \textit{NuSTAR} observations (ObsIDs 30002033006, 30002033002, 30002034004) for each epoch were reduced using the standard pipeline $\nustardas$~included in the $\heasoft~v2.26.2$. Source spectra were created using a circular region with a radius of $\sim$70$\arcsec$ and the corresponding background was estimated from a larger source-free region of the same detector. All products were extracted from the cleaned event files using XSELECT. Again, the spectra were rebinned to have minimum of 50 counts per bin. The NuSTAR spectra used in this work are modeled over the 
$3.0-30\,$keV range.      

\subsubsection{HST}
Holmberg IX X-1 has been reported as one of the ULXs with a unique optical counterpart \citep{Kaaret+04,Tao+11,Gladstone+13}. In this work, the 
intrinsic vega magnitudes were taken directly from \citet{Dudik+16} where they presented the multi-band photometric properties from the analysis 
of archival imaging data taken by the \textit{Hubble} Space Telescope (HST). A summary of archival HST data used for the analysis is presented in 
Table \ref{tab:HST-data}.

\begin{figure}
    \centering
    \includegraphics[width=0.42\textwidth]{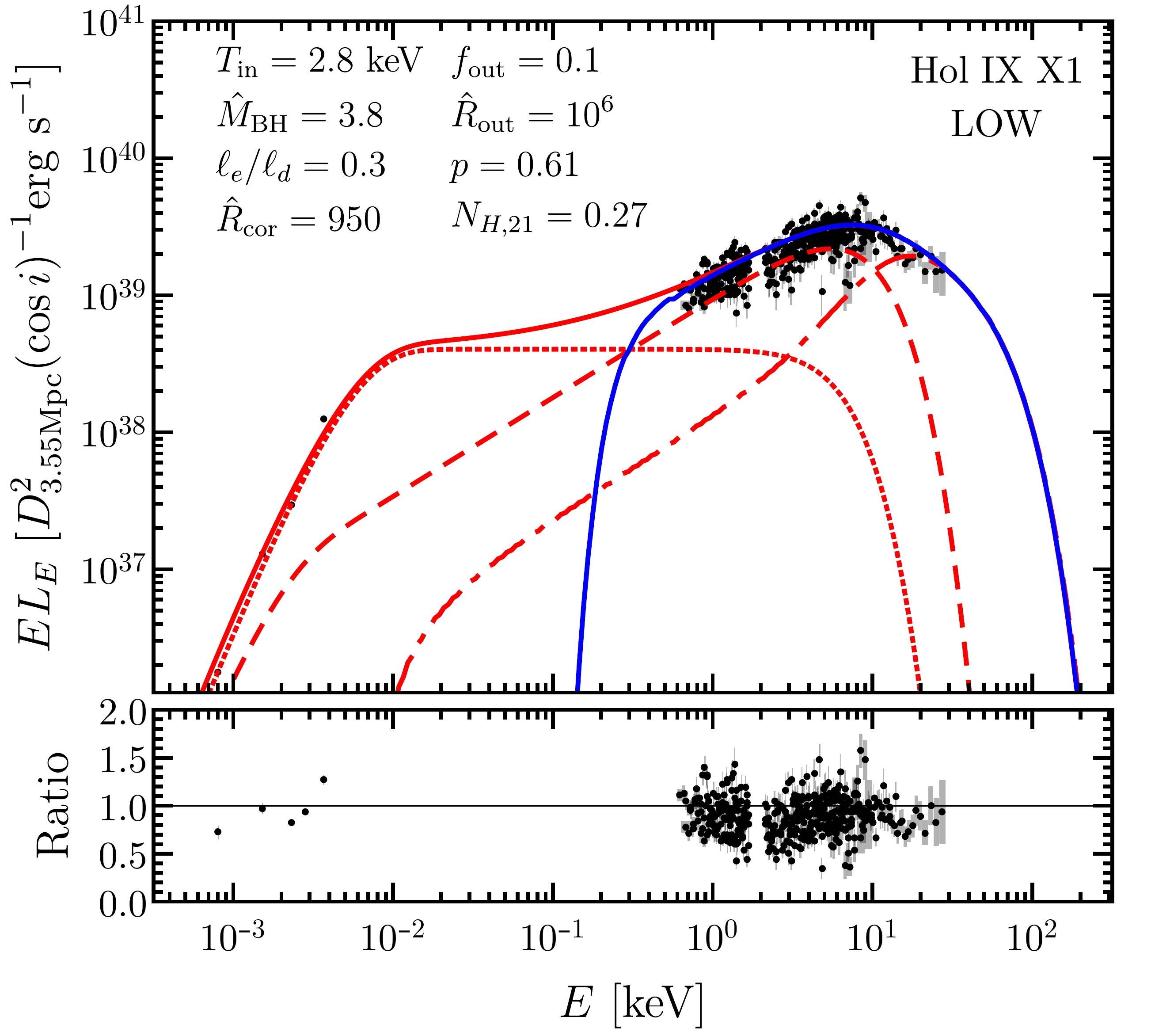}
    \includegraphics[width=0.42\textwidth]{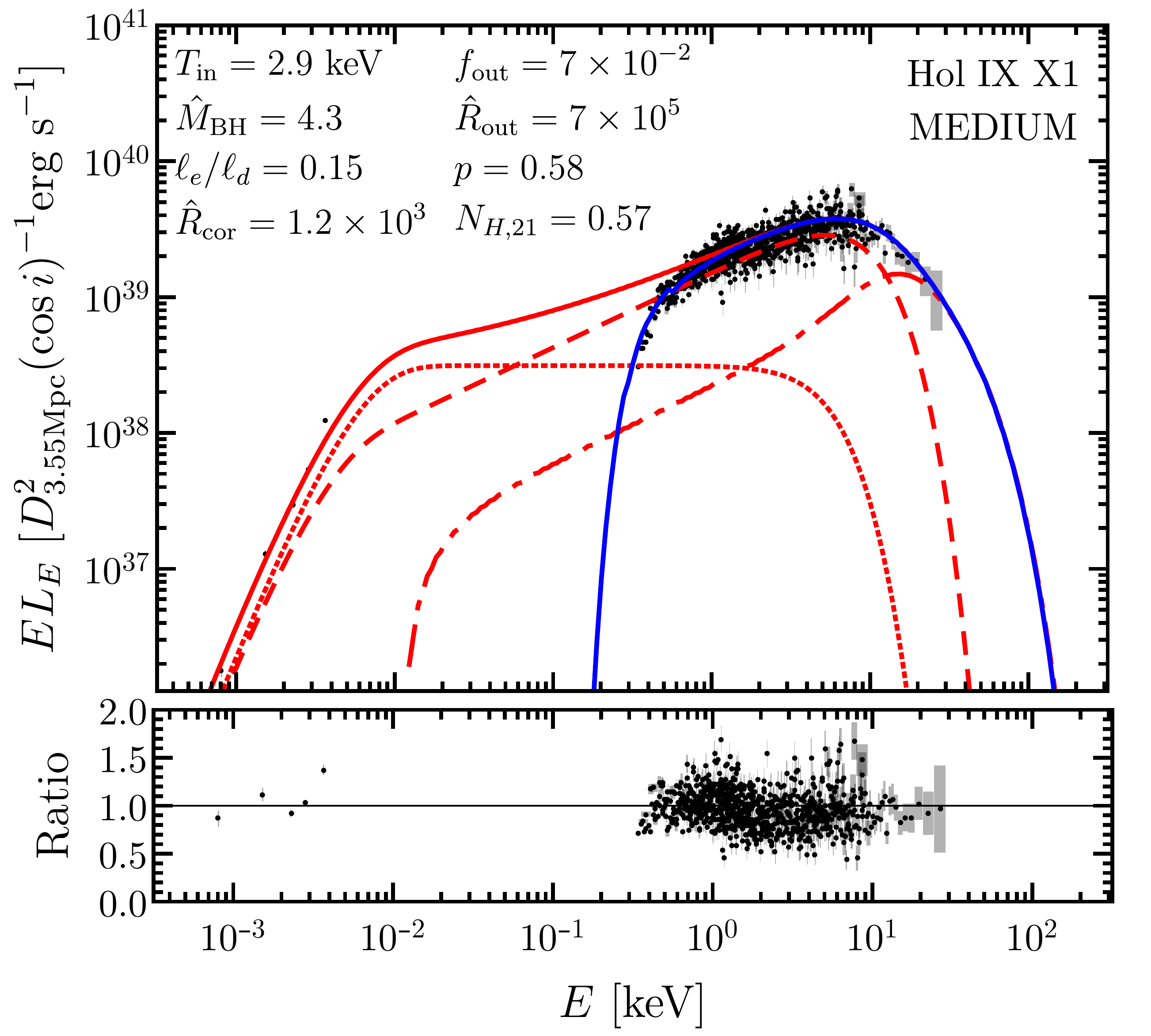}
    \includegraphics[width=0.42\textwidth]{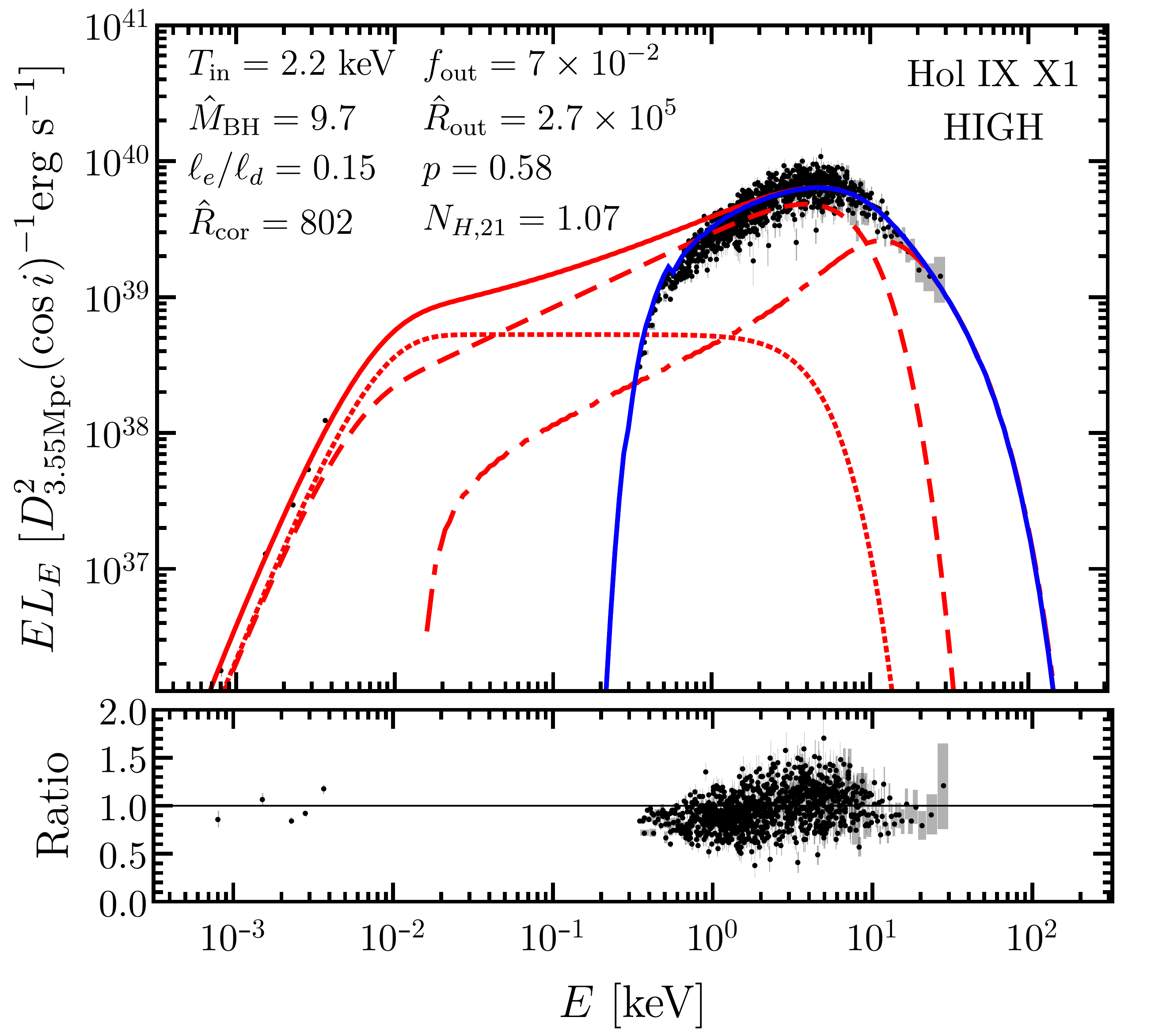}
    \caption{Two-component disk-corona model fit to the broadband spectrum of Holmberg IX X-1, shown for three different spectral states (corresponding to the 
    relative level of flux) dubbed HIGH, MEDIUM, and LOW. The data has been corrected for Galactic absorption. The solid red curve shows the de-absorbed 
    best-fit spectrum obtained with non-linear least squares and the dashed, dot-dashed, and dotted red curves show the multi-color disk, Comptonized, 
    and irradiated disk spectral components, respectively. The spectrum modified by absorption local to the source is shown by the blue curve.  
    Bottom panel shows the data/model ratio. In all cases, the steady-state corona is optically thick with fit results given in Table~\ref{tab:Best-fit-params}.}
    \label{fig:Hol-IX-X1-fit}
\end{figure}

\begin{table*}
\centering
    \begin{tabular}{|c|c|c|c|c|c|c|}
    \hline
Instrument & Obs ID & Filter & Obs. Date & Flux density & Abs. & Corrected \\
& & & & $\mu$Jy & A$_{\lambda}$ & $\mu$Jy \\
\hline
Hubble ACS  &  	GO-9796	 & F814W/I & 	2004 Feb 7 & 	2.99$\pm$0.19 & 0.47 & 	4.67$\pm$0.29 \\
Hubble ACS  &  	GO-9796	 & F555W/V & 	2004 Feb 7 & 	3.29$\pm$0.15 & 0.81 & 	7.06$\pm$0.31 \\
Hubble ACS  &  	GO-9796	 & F435W/B & 	2004 Feb 7 & 	3.87$\pm$0.10 & 1.08 & 	10.4$\pm$0.29 \\
Hubble ACS  &  	GO-9796	 & F330W/U & 	2004 Feb 7 & 	5.47$\pm$0.23 & 1.33 & 	18.6$\pm$0.74 \\
Hubble WFC3/IR  &  GO-12747  &  F160W/H & 2012 Sep 25 & 1.01$\pm$0.09 & 0.202 & 1.21$\pm$0.12 \\
\hline
\end{tabular}
\caption{A summary of archival HST data (reproduced from \citet{Dudik+16})}
\label{tab:HST-data}
\end{table*}

\begin{table*}
    \centering
    \begin{tabular}{|c|c|c|c|c|c|c|}
    \hline
    & \multicolumn{6}{c}{Model Parameters From MCMC} \\
    \hline
    Spectral State & $k_BT_{\rm in}\,$(keV) & $\hat M_{\rm BH}$ & $\ell_e/\ell_d$ & $\log_{10}\hat R_{\rm cor}$ & $p$ & $N_{H,21}$ \\
    \hline
    LOW & $2.75^{+0.59}_{-0.61}$ & $3.91^{+0.58}_{-0.29}$ & $0.20^{+0.19}_{-0.13}$ & $2.99^{+0.40}_{-0.45}$ & $0.62^{+0.10}_{-0.09}$ & 
    $0.15^{+0.37}_{-0.11}$ \\
    MEDIUM & $2.99^{+0.54}_{-0.35}$ & $4.36^{+1.71}_{-1.52}$ & $0.20^{+0.22}_{-0.13}$ & $3.00^{+0.68}_{-0.72}$ & $0.63^{+0.09}_{-0.11}$ & 
    $0.66^{+0.78}_{-0.36}$ \\
    HIGH & $2.24^{+1.09}_{-0.83}$ & $9.68^{+0.53}_{-0.60}$ & $0.23^{+0.19}_{-0.16}$ & $2.98^{+0.61}_{-0.59}$ & $0.60^{+0.11}_{-0.07}$ & 
    $1.18^{+0.89}_{-0.56}$ \\
    \hline
    \end{tabular}
    \caption{Fit parameters and corresponding 1-$\sigma$ uncertainties obtained from the MCMC simulation.}
    \label{tab:MCMC-fit-params}
\end{table*}

\begin{table*}
    \centering
    \begin{tabular}{|c|c|c|c|c|c|c|c|c|c|c|c|c|}
    \hline
    & \multicolumn{8}{c}{Best-Fit Model Parameters (8)} & \multicolumn{4}{c}{Steady-State Corona Properties (4)} \\
    \hline
    Spectral State & $k_BT_{\rm in}\,$(keV) & $\hat M_{\rm BH}$ & $\ell_e/\ell_d$ & $\hat R_{\rm cor}$ & $p$ & $N_{H,21}$  & 
    $f_{\rm out}$ & $\hat R_{\rm out,5}$ & $k_BT_e\,$(keV) & $\tau_T$ & $y_C$ & $L_c/L_d$\\
    \hline
    LOW & 2.8 & 3.81 & 0.3 & 950 & 0.61 & 0.27 & 0.1 & 10.0 & 15.5 & 4.4 & 2.4 & 0.94\\
    MEDIUM & 2.9 & 4.3 & 0.15 & 1198 & 0.58 & 0.57 & 0.07 & 7.0 & 13.3 & 3.1 & 1.1 & 0.47\\
    HIGH & 2.2 & 9.7 & 0.15 & 802 & 0.58 & 1.07 & 0.07 & 2.7 & 12.6 & 3.3 & 1.1 & 0.47\\
    \hline
    \hline
    & \multicolumn{2}{c}{$L_{\rm Bol}[{\rm erg\,s}^{-1}]$} & \multicolumn{2}{c}{$L_{(0.3-10)\,{\rm keV}}[{\rm erg\,s}^{-1}]$} & 
    \multicolumn{2}{c}{$\chi^2/{\rm dof}$} & & & & \\
    \hline
    LOW & \multicolumn{2}{c}{$1.1\times10^{40}$} & \multicolumn{2}{c}{$6.9\times10^{39}$} & \multicolumn{2}{c}{1937/433} & & & & \\
    MEDIUM & \multicolumn{2}{c}{$1.2\times10^{40}$} & \multicolumn{2}{c}{$8.4\times10^{39}$} & \multicolumn{2}{c}{4066/820} & & & & \\
    HIGH & \multicolumn{2}{c}{$2.4\times10^{40}$} & \multicolumn{2}{c}{$1.6\times10^{40}$} & \multicolumn{2}{c}{4308/970} & & & & \\
    \hline
    \end{tabular}
    \caption{Best fit parameters obtained with non-linear least squares for the three spectral states. There are eight model parameters. 
    The steady state description of the $e^\pm$-pair corona is given by the four parameters that were obtained at the end of the simulation. 
    The bolometric and $(0.3-10)\,$keV luminosities for the three spectral states are also listed. Here we use the notation 
    where $Q_x$ denotes the quantity $Q$ in units of $10^x$ times its (cgs) units.}
    \label{tab:Best-fit-params}
\end{table*}

\subsection{Spectral Analysis: BH Mass Fits}
We use the two-component multi-color disk-corona spectral model to fit the broadband spectrum of Holmberg IX X-1. The fits were first 
carried out using Markov Chain Monte Carlo (MCMC) simulations to obtain the most probable model parameters and their corresponding 1-$\sigma$ 
uncertainties. The results of the MCMC simulation are tabulated in Table \ref{tab:MCMC-fit-params} and the posterior probability distributions 
of the model parameters are shown in Appendix \ref{sec:MCMC-fits}. The best-fit model parameters were obtained from a least-squares fit and 
they are tabulated in Table \ref{tab:Best-fit-params} and the fits to the spectra are shown in Fig.~\ref{fig:Hol-IX-X1-fit}.

We find significant 
local absorption with $N_H\sim(0.3-1.0)\times10^{21}\rm{cm}^{-2}$, which is a factor $\sim(0.4-1.2)$ of the Galactic absorption in the direction 
of Holmberg IX X-1. The measurements presented in Fig.~\ref{fig:Hol-IX-X1-fit} are corrected for the Galactic absorption that contributes a 
column density of $N_H=0.813\times10^{21}{\rm cm}^{-2}$. The amount of absorption caused by material external to our Galaxy is consistent 
with that found in other works \citep[e.g.,][]{Walton+14,Luangtip+16,Walton+17}. The absorbed spectrum is shown with a blue line, for which 
we use the absorption cross-section of \citet{Wilms+00} (see their Fig.~1), and the de-absorbed spectrum is shown with a red line.

The best-fit solutions presented in Fig.~\ref{fig:Hol-IX-X1-fit} for the three different spectral states describe an accreting system 
where the mass of the accretor is $4\lesssim\hat M_{\rm BH}\lesssim10$ and the inner disk temperature is consistently high with 
$2\,{\rm keV}\lesssim k_BT_{\rm in}\lesssim3\,{\rm keV}$. Both of these parameters determine the normalization of the spectrum 
originating in the disk as well as the properties of the Comptonized spectral component. It is understood that $\hat M_{\rm BH}$ cannot change 
in between different spectral states and must remain constant. In the spectral fits it was left as a free parameter so as not to bias it 
towards any given solution. Instead, it is clear that $\hat M_{\rm BH}$ is indeed constrained in a narrow range and all spectral states are 
explained by a stellar mass BH accretor. The inner disk temperature obtained here is consistent with that found by other works 
\citep[e.g.,][]{Luangtip+16,Walton+17,Walton+18} that modeled the spectrum using two thermal spectral components, namely a combination of 
DISKBB and DISKPBB models. The DISKPBB component was always found to be hotter of the two and that is what the MCD spectral component 
(dashed curve) shown in Fig.~\ref{fig:Hol-IX-X1-fit} represents.

The hard excess above $10\,$keV, as revealed by \textit{NuSTAR}, is explained here by the Comptonized emission (dash-dotted curve) originating in 
the corona. In empirical model spectral fits this component is typically explained using a CUTOFFPL or SIMPL model. We obtain the Comptonized 
emission by injecting into the corona a non-relativistic and monoenergetic $e^\pm$-pair distribution with $\gamma_{\rm inj}=1.1$ that corresponds 
to an effective temperature of $k_BT_{\rm eff}=(2/3)(\langle\gamma_e\rangle-1)m_ec^2\approx34\,$keV. Even though the pairs are being continuously 
injected at this temperature, the mean temperature of the pair distribution in the corona declines rapidly due to the large compactness ($\ell_d\sim60$) 
of the disk emission that Compton cools the pairs to a lower temperature as the system approaches a steady state. The steady state distribution of 
pairs is a Maxwellian with temperature in the range $k_BT_e \sim (12-15)\,$keV and the Thomson optical depth of the corona is modest ($\tau_T\sim3.5$). 
The Compton-$y$ parameter of the coronal plasma is of order unity and the system never reaches the stage where saturated Comptonization would establish 
a pronounced Wien peak at $E=3k_BT_e$, the energy beyond which the model spectrum would drop off exponentially.

When comparing the model parameters for the three states, the inner disk temperature, $T_{\rm in}$, shows a small spread which explains the narrow 
spread in the spectral peak energy. Since the disk luminosity scales as $L_d\propto T_{\rm in}^4$, even a small change in the inner disk temperature can 
bring a large change in the normalization of the peak luminosity. However, the trend appears to be reversed when comparing the MEDIUM and HIGH spectral 
fits, where $T_{\rm in}$ is larger in the MEDIUM spectral state. The rise in flux density in the HIGH state is accounted for by the larger value for 
$\hat M_{\rm BH}$. This negative trend between disk luminosity and inner disk temperature was also noted by \citet{Luangtip+16}. Indeed, it finds a natural 
explanation in super-critical accretion models \citep[see, e.g., extensive discussion in][]{Luangtip+16,Walton+17}.

The disk temperature profile power law index remains almost constant, with $p\approx0.6$, among all the spectral states. This result is also 
consistent with other works \citep[e.g.,][]{Walton+17}.

The properties of the coronal plasma are rather uniform among the three states, which is indicative of the fact that it is mainly the disk emission 
that is changing. The coronal plasma is not photon starved (i.e. $\ell_e/\ell_d < 1$) which explains why the steady state particle temperature ($\sim14\,$keV) 
is lower than their injection temperature ($34\,$keV), where the particles have been cooled due to Comptonization. The size of the corona also remains 
more or less constant between the different flux states, with $\hat R_{\rm cor}\sim10^3$.

A large spread is seen in the value of the absorption column density local to the source. This affects the degree of X-ray flux suppression at 
energies $\lesssim2\,$keV, which appears to be significant for the HIGH state. Since there is insufficient data below $0.5\,$keV for the LOW state, the 
inferred value of the column density cannot be trusted. This is also clear from the MCMC fit shown in Fig.~\ref{fig:MCMC-fits} where a distinct peak 
in the posterior distribution away from the boundary values is missing.

To constrain the size of the disk and the disk irradiation physics we have included archival HST observations and carried out broadband spectral 
fits. These observations are not contemporaneous with the X-ray observations and so do not probe the effect of spectral states on the inferred disk 
irradiation model parameters directly. However, such an exercise does demonstrate the potential of fitting contemporaneous broadband observations with 
our model. The inferred values of the two parameters $f_{\rm out}\sim0.08$ and $\hat R_{\rm out}\sim7\times10^5$ are consistent with other works that 
fit broadband optical/UV and X-ray data of ULXs \citep[e.g.,][]{Gierlinski+09,Sutton+14}. The irradiated disk model fit to the HST data depends on the 
model fits to the X-ray data but not vice versa. Therefore, even though the HST observations included here are not contemporaneous, these do not affect 
the other six model parameters, including $\hat M_{\rm BH}$, that rely solely on the X-ray data. This is the reason behind not including the HST data in 
the MCMC fits. Furthermore, due to insufficient UV data that would constrain the spectral break, the irradiated disk model is degenerate in the two parameters 
and as such $f_{\rm out}$ and $\hat R_{\rm out}$ cannot be constrained well. Therefore, the obtained irradiated disk parameters are not unique.

\section{Discussion}\label{sec:discussion}
In this work we have tried to understand the origin of what appears to be a three-component X-ray spectrum of ULXs in the $(0.3-30)\,$ keV band using a 
physical model that features a multi-color disk plus Comptonizing corona. Here the inner accretion disk contributes a MCD thermal spectrum that 
describes the X-ray spectrum below $10\,$keV whereas a hotter corona of $e^\pm$-pairs Comptonizes the disk radiation to give the high-energy 
component above $10\,$keV. We have used a time-dependent kinetic code to evolve the coupled lepto-photonic equations of radiation transfer to obtain 
the two-component X-ray spectrum. In addition, we self-consistently produce the irradiated disk-component that explains the low-energy optical/UV spectrum 
of ULXs.

The prime motivations behind this work were to (i) explain as self-consistently as possible the origin of the X-ray spectrum and (ii) constrain the 
mass of the central compact object -- a major open question, which is in fact one of the parameters of our model. Here we apply our spectral model to fit to the 
different spectral states of one of the most well studied and luminous ULXs, Holmberg IX X-1, and constrain the mass of the accretor. Our working assumption 
is that the accretor is a BH and we constrain its mass to be in the range $4\lesssim\hat M_{\rm BH}\lesssim10$. 
Here $\hat M_{\rm BH}\equiv \alpha M_{\rm BH}/M_\odot$ which depends on the BH spin and therefore the mass only goes up for a spinning BH for which 
$1/6\leq\alpha<1$. The normalization of the 
MCD component scales as $L_d\propto M_{\rm BH}^2T_{\rm in}^4$, and therefore a lower $\hat M_{\rm BH}$ naturally means a higher inner disk temperature
$T_{\rm in}$ and vice versa. In addition, since the luminosity changes with source distance and inclination, such that 
$L\propto D^2/\cos i$, Eq.~(\ref{eq:L_d}) yields the scaling $M_{\rm BH}\propto D/\sqrt{\cos i}$. Therefore, a larger distance or larger inclination angle 
would tend to increase the $\hat M_{\rm BH}$ estimate. A similar change results in $\hat R_{\rm cor}$ when $T_{\rm in}$ can be fixed since 
$\ell_d\propto\hat M_{\rm BH}/\hat R_{\rm cor}$ (when $\hat R_{\rm cor}>1$) and to obtain the same spectrum, albeit with a different overall normalization, 
$\ell_d$ must remain unchanged.

Previous BH mass estimates in Holmberg IX X-1 have ranged from $(10-10^3)M_\odot$ based on spectral modeling \citep[e.g.,][]{Wang-02,Miller+03,Kong+10} 
and $(50-200)M_\odot$ from the detection of quasi-periodic oscillations \citep[QPOs;][]{Dewangan+06}. In general, mass estimates in ULXs have relied on a 
few different methods 
\citep{Miller+03,Miller+04,Miller+13}. For example, the standard thin-disk scaling relation $T_{\rm in}\propto M_{\rm BH}^{-1/4}$ is often employed with 
the underlying assumption that ULXs are scaled up versions of known accreting sources. Noting that $k_BT_{\rm in,GBHB}\sim1\,$keV in Galactic BHBs, that 
are powered by stellar-mass BHs with $M_{\rm BH}\sim10M_\odot$, as compared to $k_BT_{\rm in,ULX}\sim0.1-0.3\,$keV, the temperature of the cooler 
thermal disk emission, this yields $M_{\rm ULX}\sim10^3-10^5M_\odot$. Accordingly, the higher disk temperatures obtained from the spectral fits in this work 
are commensurate with having a stellar mass BH in Holmberg IX X-1. Alternatively, when using the MCD model a more direct mass estimate is possible where the 
normalization of the disk luminosity depends on 
$M_{\rm BH}$. This of course makes the assumption that the disk extends all the way to $R_{\rm ISCO}$. Indeed, having made the same assumptions in this work, 
we were able to constrain $M_{\rm BH}$ directly as it is one of the model parameters. Another useful way of constraining the mass is by comparing the bolometric 
luminosity with $L_{\rm Edd}$ assuming an accretion efficiency $\eta\equiv L_{\rm Bol}/L_{\rm Edd}<1$. Having determined the masses using an independent 
method, we can use $L_{\rm Edd}$ to constrain $\eta$. For the obtained mass range the Eddington luminosity is 
$L_{\rm Edd}\simeq(0.5-1.3)\times10^{39}\alpha^{-1}\,{\rm erg\,s}^{-1}$. By comparing it with the bolometric luminosities obtained for the three flux states, 
we find that $\eta\sim20\alpha$, which indicates super-critical accretion.

While low ($k_BT_{\rm in}\sim0.1-0.3\,$keV) disk temperatures are the hallmark of ULXs, some works do find high ($k_BT_{\rm in}\gtrsim1\,$keV) disk temperatures, 
particularly for the highly luminous states, when fitting in many cases with dual thermal models that feature either two MCD or MCD plus blackbody components 
\citep{Stobbart+06,Walton+14,Luangtip+16,Walton+17}. As argued above, in the standard thin-accretion disk models lower $T_{\rm in}$ necessarily yields 
$M_{\rm BH}>100M_\odot$, one promising way to obtain lower BH masses is by having higher $T_{\rm in}$. In that case, we find that as $T_{\rm in}$ is raised, 
moving the thermal peak to higher energies, the spectral slope below the peak cannot be explained with the disk temperature profile having power-law index $p=0.75$. 
Instead, the temperature profile must deviate from the standard thin-disk relation and must be harder with $p\sim0.6$, a result consistent with other 
works \citep[e.g.][]{Walton+14,Walton+17}. Such a change might be indicative of radiation pressure effects operating in the hot inner disk, as would be 
obtained in super-critical accretion, which would alter the temperature profile, as in a slim accretion disk \citep[e.g.][]{Abramowicz+88,Poutanen+07}.

The spectral model employed in this work has only two components in the X-ray band (the third component being the irradiated disk component that 
appears only in the optical/UV energy band), in contrast to many works that use two spectral components to fit the softer X-ray emission below $\sim10\,$keV and 
an additional component to fit the hard excess revealed by \textit{NuSTAR} above $\sim10\,$keV. As such, we only model the hotter of the two thermal 
components with an MCD for the softer X-ray emission. The harder excess in our model is given by the Comptonized emission, which is typically modeled using 
a CUTOFFPL or SIMPL emperical model in many works. Although the cooler of the two thermal components contributing to the softer X-ray emission is found 
to be subdominant \citep[e.g.,][]{Luangtip+16,Walton+17}, its inclusion is statistically demanded. The lack of such a component in the present 
work leads to some structure in the data/model ratios, as shown in Fig.~\ref{fig:Hol-IX-X1-fit}, and produces rather high values of the $\chi^2$ fit statistic.

Many works model the hotter thermal component as Comptonized emission by thermal electrons using either COMPTT or EQPAIR models 
\citep[e.g.,][]{Stobbart+06,Gladstone+09,Luangtip+16}. In such a case the softer thermal component would be the MCD emission. When we try to model 
the X-ray spectrum of Holmberg IX X-1 with our model in this manner, we find that a lower $T_{\rm in}$ can only be accommodated if 
$\hat R_{\rm cor}\ll1$. This further implies that the solid angle offered by the corona is rather small and most of the disk emission is not 
intercepted. Therefore, the Comptonized emission should be sub-dominant and ultimately inadequate to explain the X-ray spectrum at energies 
beyond $\sim1\,$keV. As a result, the cooler of the two thermal components, when statistically required, must have a different origin in our 
model. Its origin and significance will be investigated in future works.

The hotter thermal component finds a natural explanation in models featuring super-critical accretion with massive outflowing winds 
\citep{Abramowicz+88,Poutanen+07,King-09,Dotan-Shaviv-11,Ohsuga-Mineshige-11,Takeuchi+13}. Radiation pressure effects lead to large 
scale heights of the disk and massive winds that are shaped into a funnel geometry around the rotational axis of the compact source. The softer thermal 
component then emerges as photospheric emission from the walls of this funnel while the hotter component has it origin in the inner regions of the 
accretion flow. Then, for a given line-of-sight the different spectral states can occur due to changes in the accretion rate which would alter the 
scale height, and correspondingly the position of the funnel wall, of the accretion disk \citep[][]{Middleton+15}. Alternatively, if the accretion disk 
precesses, that too can lead to spectral changes due to the change in the inclination angle of the line-of-sight \citep[e.g.][]{Luangtip+16}.

In this work, we have also attempted to self-consistently explain the optical/UV emission using the irradiated disk model \citep{Gierlinski+09}. The 
self-consistency comes from obtaining the ratio $L_c/L_d$ from the steady state solution. The value of this parameter is usually assumed 
a priori in all works. When the optical/UV data is missing the spectral break or shoulder (around $\sim10\,$eV), as was the case in the archival 
HST data used here, the two parameters, $\hat f_{\rm out}$ and $\hat R_{\rm out}$, become degenerate that precludes obtaining strong constraints. 
Future nearly contemporaneous observations by, e.g. \textit{Swift/UVOT} and \textit{XRT}, where the former might be able to localize the spectral break in 
some sources, with \textit{NuSTAR} are encouraged.

\subsection{Caveats: Super-Critical Accretion}
As shown earlier, for the narrow range of BH masses obtained in this work, we find accretion to be proceeding in the super-critical regime in Holmberg 
IX X-1. This naturally questions the validity of using a $p$-free thin accretion disk formalism, as the emission might be arising from or significantly 
modified by massive radiative winds that produce flux enhancements due to geometric beaming and also impose inclination angle dependent spectral variations. 
The bolometric isotropic-equivalent luminosity of a super-critically accreting BH is \citep{Begelman+06,Poutanen+07,King-09}
\begin{equation}
    L_{\rm Sph}\simeq L_{\rm Edd}\frac{(1+\ln\dot m)}{b}=2L_d\simeq2\eta L_{\rm Edd}\,,
\end{equation}
where $\dot m=\dot M/\dot M_{\rm Edd}$ is the Eddington normalized mass accretion rate and $b$ is the geometric beaming factor. For $\eta\sim20$ and 
if $b\gtrsim0.1$ then this yields $\dot m\gtrsim20$. If geometric beaming is indeed important in Holmberg IX X-1, then the intrinsic flux normalization 
is lower by a factor of $0.1\lesssim b\lesssim1$, which implies that the true $\hat M_{\rm BH}$ might be smaller by a factor of $0.3\lesssim\sqrt b\lesssim1$.

In addition, the one-zone model used in this work is incapable of addressing the radiation transfer physics and accounting for 
the flow geometry. Indeed, this is a limitation of our model and caveat for the obtained results. In order to account for the different 
effects in such a scenario, a more sophisticated radiation transfer model \citep[e.g.][]{Kawashima+12} that also accounts for spectral 
changes due to different inclinations is more appropriate.

The complex geometry of the outflow may also affect the disk irradiation physics. In a sub-Eddington system, which lacks such outflows, the corona 
irradiates the outer disk directly. In contrast, the massive outflows in a super-critical system are optically thick at large inclinations, and 
as a result, the coronal emission may be highly suppressed. Therefore, a proper radiation transfer treatment is again required to fully understand 
these effects.

\section{Conclusions}\label{sec:conclusions}

Our broadband spectral modeling, using a self-consistent two-component MCD plus Comptonizing corona model for the X-ray emission, 
suggests that Holmberg IX X-1 harbors a stellar mass BH with mass $4\lesssim\hat M_{\rm BH}\lesssim10$. With the discovery of pulsations in at least 
six ULXs, the consensus has now shifted to most of these systems likely hosting accreting NSs and in some cases stellar mass BHs. No pulsations have 
been detected from Holmberg IX X-1 thus far, which does favor the conclusions reached in this work. However, non-detection of pulsations is not 
a sufficient condition to rule out a NS host. Finally, this work demonstrates the use of physical models to arrive at more self-consistent 
spectral fit solutions. It is envisaged that this technique, when applied to a larger sample of ULXs, will yield important clues to their nature.
\\
\section*{Acknowledgements}
\noindent
We are grateful to the anonymous referee for several valuable comments that significantly improved the quality of this work.
RG is supported by the ISF-NSFC joint research program (grant No. 3296/19). 
This research is supported by the Scientific and Technological Research Council of Turkey (TUBITAK) through project  number 119F334.

\section*{Data Availability}
The data underlying this article will be shared on reasonable request to the corresponding author.

\bibliographystyle{mnras}
\bibliography{refs.bib}

\begin{appendix}

\section{Model Comparison}\label{sec:Model-Comparison}

\begin{figure}
    \centering
    \includegraphics[width=0.48\textwidth]{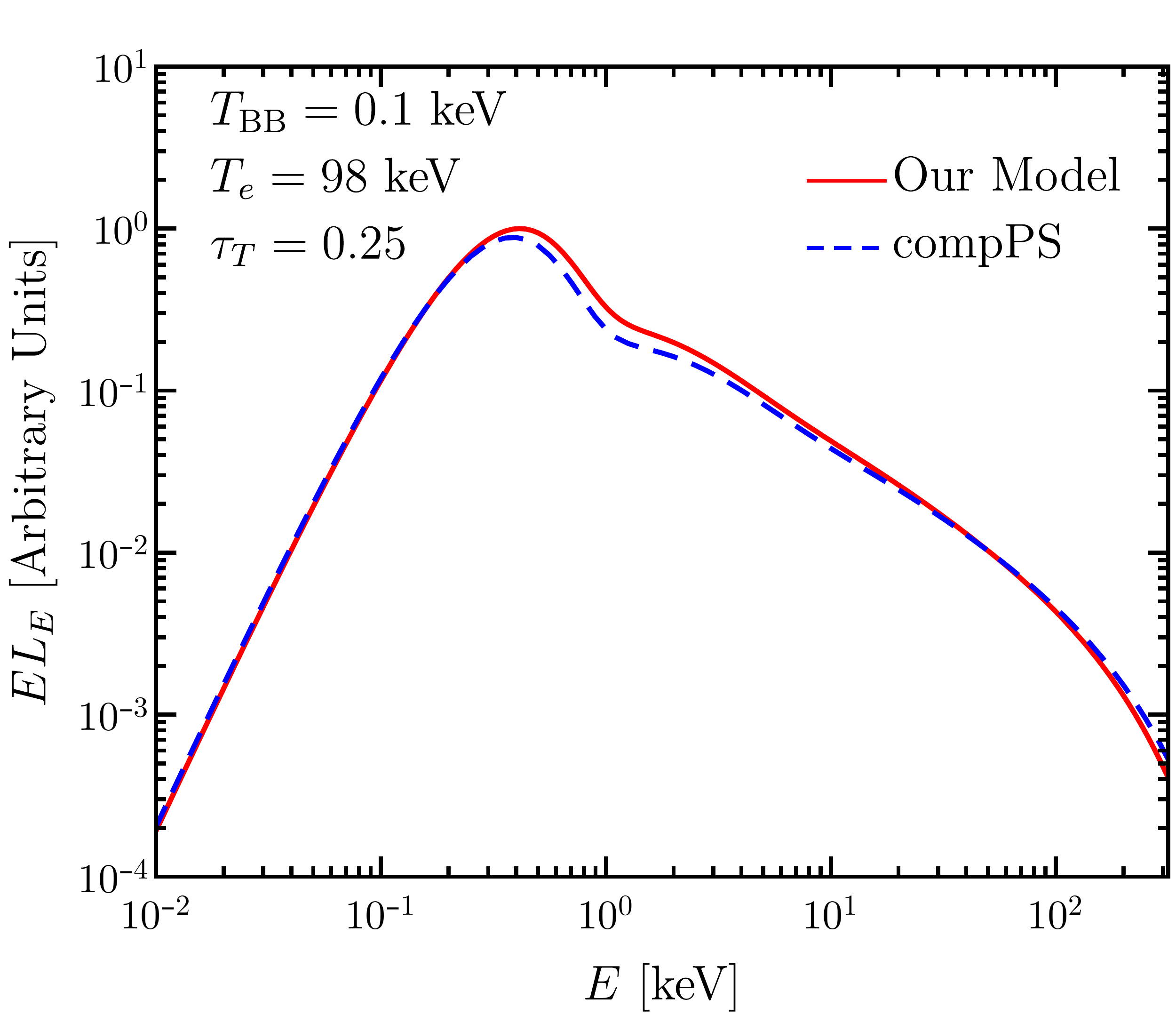}
    \caption{Model comparison with compPS using a simple thermal Comptonization setup.}
    \label{fig:compPS-comparison}
\end{figure}

Here we compare the thermally Comptonized spectrum from our model with CompPS \citep{Poutanen-Svensson-96}, a popular Comptonization model from the 
literature that also uses the exact Compton scattering kernel producing more accurate spectra. CompPS uses blackbody (at temperature $T_{\rm BB}$) or MCD 
emission for the seed photon spectrum and calculates the Comptonized emission for a thermal (at temperature $T_e$), power-law, or hybrid particle distribution. 
In that regard, the capabilities of our time-dependent numerical code are similar to CompPS, but the latter makes the steady state assumption that allows 
it to be computationally much less expensive. We compare the results from the two models in Fig.~\ref{fig:compPS-comparison} for Thomson optical depth 
$\tau_T$ of the electrons and show that our model achieves good agreement with compPS.

\section{MCMC Model Fits}\label{sec:MCMC-fits}

\begin{figure}
    \centering
    \includegraphics[width=0.44\textwidth]{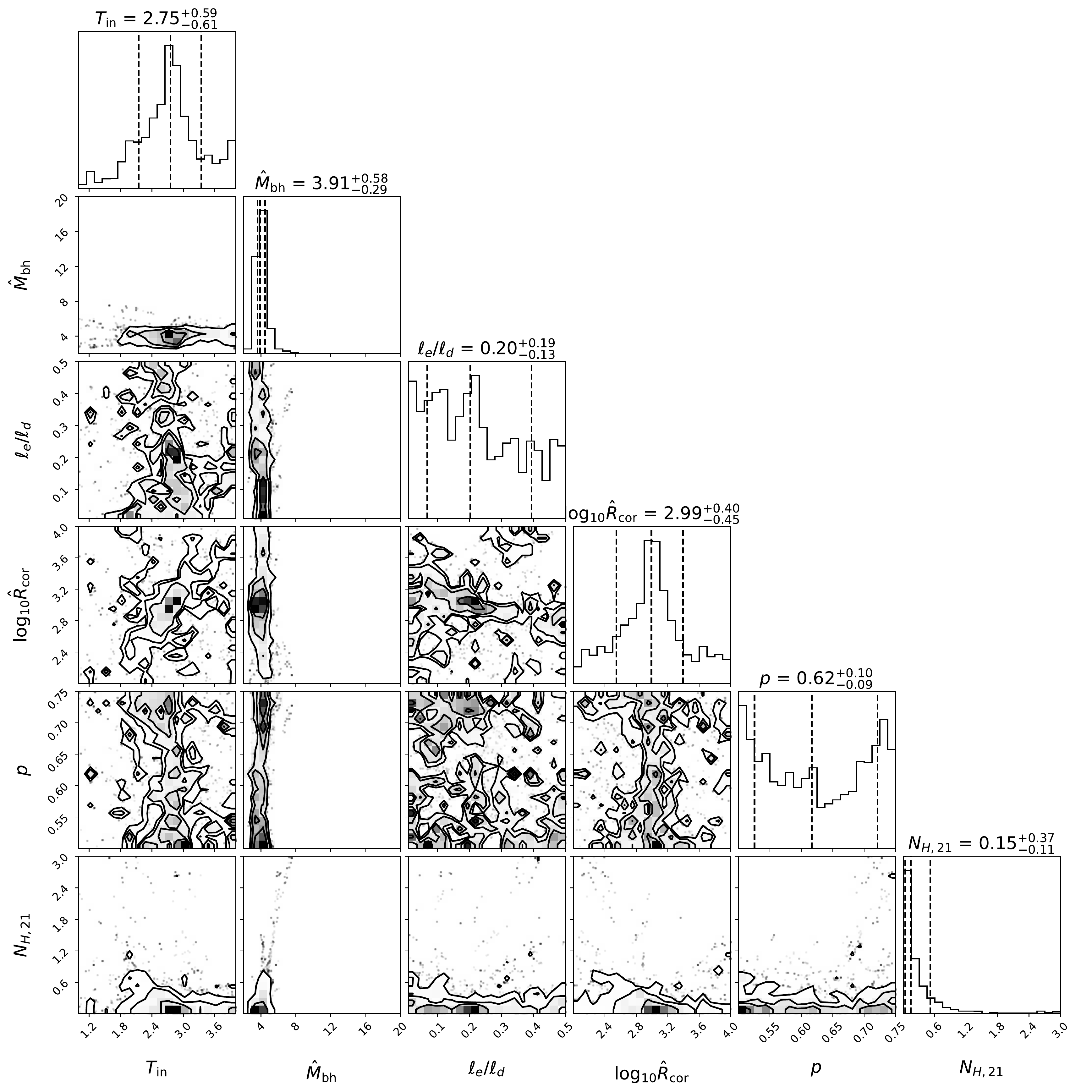} \\
    \includegraphics[width=0.44\textwidth]{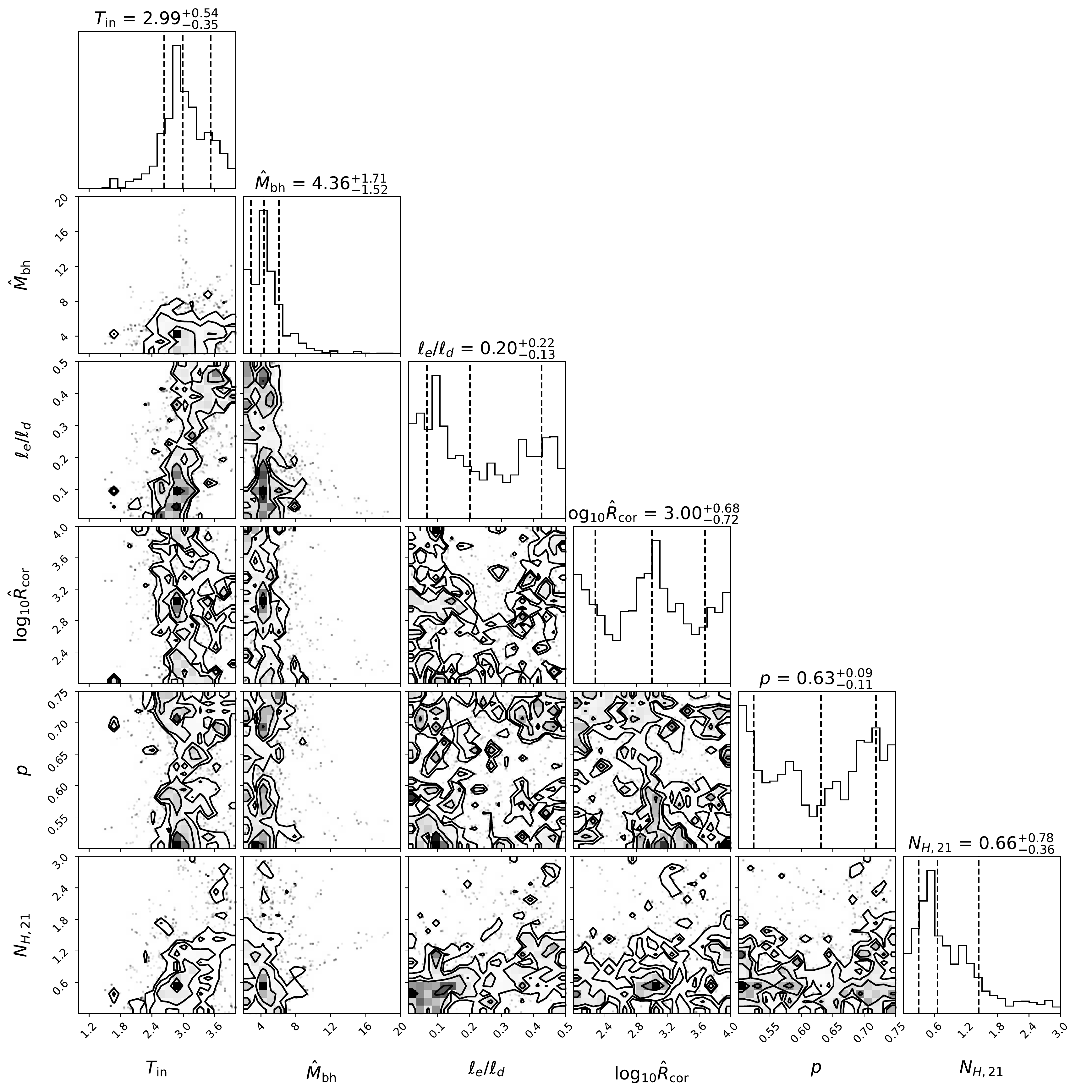} \\
    \includegraphics[width=0.44\textwidth]{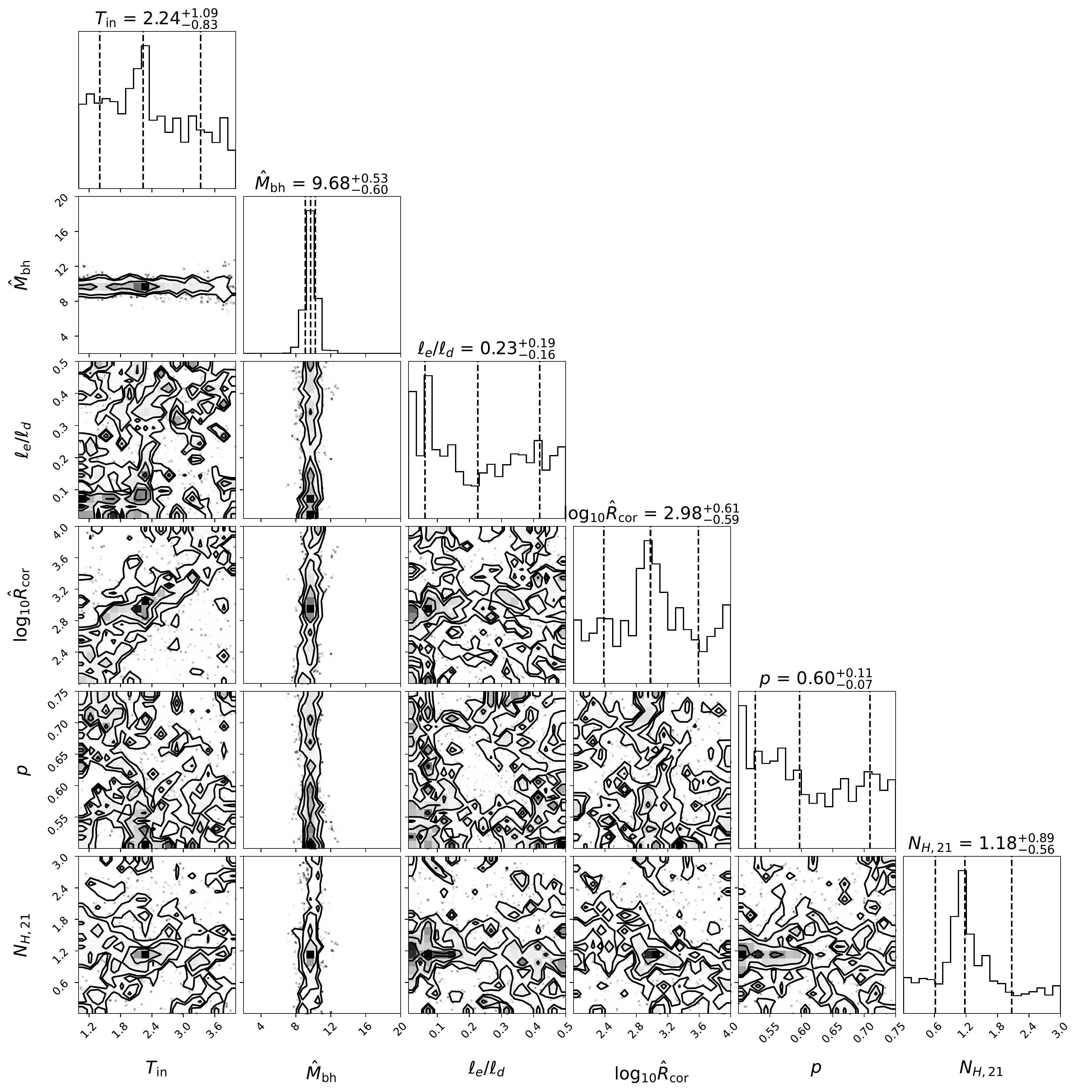}
    \caption{MCMC fit to the LOW (top), MEDIUM (middle), and HIGH (bottom) states. The posterior distributions are 
    obtained with $N_{\rm walker}=28$ and $N_{\rm steps}=200$ for each walker with a mean acceptance fraction of 0.29.}
    \label{fig:MCMC-fits}
\end{figure}

To constrain the BH mass we carried out MCMC simulations using the python package \texttt{emcee}. The disk-corona model was used to fit only the X-ray 
data, and therefore no constraints were obtained for the $f_{\rm out}$ and $\hat R_{\rm out}$ parameters as they describe the irradiated disk 
component. These parameters are also degenerate given the sparse optical/UV data that is missing the spectral break near $\sim10\,$eV. This break 
is required to break the degeneracy.

The MCMC chains were constructed with $N_{\rm walker}=28$ and $N_{\rm steps}=200$ for each walker with a burn-in of 20 steps. 
Lowering $N_{\rm walker}$ by a factor of 2 and increasing $N_{\rm steps}$ by the same factor yields the same final result. Due to the computationally 
intensive nature of the numerical code used 
for producing the X-ray spectrum, the total number of steps had to be kept small. To ensure good sampling of the parameter distribution, a rule of 
thumb is to have a mean acceptance fraction between 0.2 and 0.5. In all the MCMC simulations we find the mean acceptance fraction to be 0.29, which 
indicates good sampling.

From the posterior distributions, we find that the inner disk temperature, BH mass, and the size of the corona are well constrained in all the 
spectral states. Since our working assumption for the compact source is a BH, the probed mass range was limited from below at $\hat M_{\rm BH}\geq2$. 
The model parameters that are constrained most poorly are $\ell_e/\ell_d$ and $p$ due to some degeneracy in how they affect the X-ray spectrum.




\end{appendix}


\bsp	
\label{lastpage}
\end{document}